\documentclass[sigconf, nonacm]{acmart}

\AtBeginDocument{%
  }

\setcopyright{acmlicensed}
\copyrightyear{2025}
\acmYear{2025}
\acmDOI{XXXXXXX.XXXXXXX}
\acmConference[Conference acronym 'XX]{Make sure to enter the correct
  conference title from your rights confirmation email}{Oct 27--31,
  2025}{Dublin}
\acmISBN{978-1-4503-XXXX-X/2018/06}
\usepackage{multirow} 
\usepackage{tabularx}

\begin{document}



\title{DualDub: Video-to-Soundtrack Generation via Joint Speech and Background Audio Synthesis}

\author{Wenjie Tian}
\email{twj@mail.nwpu.edu.cn}
\affiliation{%
  \institution{Northwestern Polytechnical University}
  \city{Xi'an}
  \country{China}
}
\author{Xinfa Zhu}
\email{xfzhu@mail.nwpu.edu.cn}
\affiliation{%
  \institution{Northwestern Polytechnical University}
  \city{Xi'an}
  \country{China}
}

\author{Haohe Liu}
\email{haohe.liu@surrey.ac.uk}
\affiliation{%
  \institution{Centre for Vision Speech and Signal Processing, University of Surrey}
  \city{Guildford}
  \country{UK}
}

\author{Zhixian Zhao}
\email{zxzhao@mail.nwpu.edu.cn}
\affiliation{%
  \institution{Northwestern Polytechnical University}
  \city{Xi'an}
  \country{China}
}

\author{Zihao Chen}
\email{chenzihao@ztgame.com}
\affiliation{%
  \institution{AI Lab, Giant Network}
  \city{Shanghai}
  \country{China}
}

\author{Chaofan Ding}
\email{dingchaofan@ztgame.com}
\affiliation{%
  \institution{AI Lab, Giant Network}
  \city{Shanghai}
  \country{China}
}

\author{Xinhan Di}
\email{dixinhan@ztgame.com}
\affiliation{%
  \institution{AI Lab, Giant Network}
  \city{Shanghai}
  \country{China}
}

\author{Junjie Zheng}
\email{zhengjunjie@ztgame.com}
\affiliation{%
  \institution{AI Lab, Giant Network}
  \city{Shanghai}
  \country{China}
}
\author{Lei Xie}
\email{lxie@nwpu.edu.cn}
\affiliation{%
  \institution{Northwestern Polytechnical University}
  \city{Xi'an}
  \country{China}
}



\begin{abstract}

While recent video-to-audio (V2A) models can generate realistic background audio from visual input, they largely overlook speech, an essential part of many video soundtracks. This paper proposes a new task, video-to-soundtrack (V2ST) generation, which aims to jointly produce synchronized background audio and speech within a unified framework. 
To tackle V2ST, we introduce DualDub, a unified framework built on a multimodal language model that integrates a multimodal encoder, a cross-modal aligner, and dual decoding heads for simultaneous background audio and speech generation. Specifically, our proposed cross-modal aligner employs causal and non-causal attention mechanisms to improve synchronization and acoustic harmony. Besides, to handle data scarcity, we design a curriculum learning strategy that progressively builds the multimodal capability. Finally, we introduce DualBench, the first benchmark for V2ST evaluation with a carefully curated test set and comprehensive metrics. 
Experimental results demonstrate that DualDub achieves state-of-the-art performance, generating high-quality and well-synchronized soundtracks with both speech and background audio.
DualBench and generated samples of DualDub are available at \url{https://anonymous.4open.science/r/DualBench-56E5}.

\end{abstract}



\maketitle

\section{Introduction}

\begin{figure}[ht]
  \centering
  \includegraphics[width=1\linewidth]{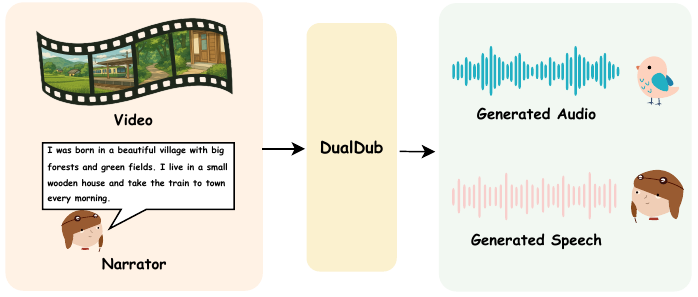}  
  \caption{Overview of the proposed DualDub. DualDub focuses on the video-to-soundtrack generation task with joint speech and background audio synthesis.}
  \label{fig_dualdub_overview}
\end{figure}


Modeling the relationship between visual input and corresponding background sounds has advanced significantly in recent work on video-to-audio generation~(V2A)~\cite{iashin2021taming-specvqgan}. Without the need for manual sound design, these systems can produce realistic audio effects corresponding to visual scenes, such as birds flying, doors closing, or waves crashing. In post-production workflows, these techniques are being increasingly utilized to enhance video content and automate Foley generation.


As a crucial element of video soundtracks and auditory context, speech is still mostly ignored in current V2A research~\cite{mmaudio, V2a-mapper, Diff-foley, FoleyCrafter, seeandhear, Frieren, Sonicvisionlm}.
While most V2A models include speech data during model training, speech-related conditions such as transcriptions or speaker prompts are not integrated into the V2A modeling. This means speech is treated as another sound effect, resulting in the generation of unintelligible and uncontrollable speech that is often rumbling-like without a clear linguistic structure.
This significantly limits the application of V2A in scenarios where both background audio and spoken content are needed.

This leads us to the question: \textit{Can we build a unified model that generates both intelligible speech and background audio in a temporally aligned and coherent way?}

Several key aspects make the generation of both background audio and speech in a unified framework challenging.
Firstly, in contrast to only generating background audio, speech generation requires more fine-grained temporal synchronization and acoustic modeling, such as prosody, duration, and speaker timbre modeling. 
These aspects are essential for producing intelligible and natural-sounding speech and are fundamentally different from the data-driven approaches used for background audio modeling.
A possible workaround is to use two separate models to generate background audio and speech independently and then combine them. However, this approach often leads to synchronization issues and disharmony between the background audio and speech, such as inconsistent volume levels, unnatural timing, and inconsistent acoustic characteristics.
Besides, high-quality datasets with paired video, audio, transcript, and speech~\cite{v2c, Animdataset} are scarce, which limits the generalization of models.
Finally, as metrics used in previous work mainly evaluate the performance of V2A, how to comprehensively and accurately evaluate generated speech and audio remains an open question.

In this work, we propose a more practical and complete task, video-to-soundtrack (V2ST) generation, aiming to generate the entire soundtrack for a video, including both background audio and dubbed speech, in a unified framework.
To address this challenging task, as shown in Figure \ref{fig_dualdub_overview}, we propose DualDub, the first unified framework for V2ST generation, and introduce DualBench, a dedicated benchmark for evaluating this task. Motivated by the strong modeling capability of language models (LMs)~\cite{lm3, lm4, audiogpt, qwen2, llama, uniaudiolm}, especially in speech generation~\cite{ CosyVoice2, megatts2, llasa, FELLE}, DualDub takes a multimodal language model as the backbone and equips a multimodal encoder and decoder. This structure enables the unified modeling of background audio and speech and better semantic consistency in speech generation. In addition, we propose a cross-modal aligner incorporating different cross-attention mechanisms to ensure synchronization and harmony of the generated soundtrack. This module leverages non-causal cross-attention for fusing visual information and causal cross-attention between audio and speech to maintain generation consistency. Moreover, to overcome the data scarcity, we introduce a curriculum learning strategy that progressively extends the multimodal capability of DualDub. This strategy initially trains DualDub on video-to-audio, then text-to-speech, and finally video-text-to-audio-speech. At each stage, corpora from the previous stage are maintained, and minimal new corpora are used to learn new capabilities, which significantly reduces data requirements while boosting data efficiency and generation quality. 

Finally, for comprehensive validation, we introduce DualBench, the first open benchmark for evaluating V2ST generation, addressing the critical need for holistic generated soundtrack assessment. DualBench includes a novel test set tailored for video-to-soundtrack tasks and comprehensive evaluation metrics covering generation quality, video synchronization, and audio-speech harmony.
Our approach is validated through extensive experiments on the DualBench. Experimental results reveal that DualDub exhibits robust performance, particularly in terms of audio and speech quality. Moreover, compared to independent generation approaches, DualDub generates highly synchronized and harmonious soundtracks for videos, marking the first attempt at the video-to-soundtrack task. The project of DualDub is available at~\url{https://anonymous.4open.science/r/DualBench-56E5}.

The key contributions of our work are summarized as follows:
\begin{itemize}
    \item We propose DualDub, the multimodal language model that enables video-to-soundtrack generation by jointly synthesizing background audio and speech.
    \item We propose a novel aligner incorporating non-causal and causal cross-attention mechanisms to improve multimodal synchronization and harmony.
    \item We introduce a novel curriculum learning strategy that effectively leverages limited multimodal data to alleviate data scarcity and enhance the generation capabilities of DualDub.
    \item We propose a new open-source benchmark, DualBench, for the systematic evaluation of the video-to-soundtrack task, including a paired multimodal test set and comprehensive metrics.
\end{itemize}

\section{Related Work}

Background audio and speech play distinct yet complementary roles in the proposed video-to-soundtrack generation task. Background audio sets the atmosphere for video scenes, while speech conveys the dialogue content and emotions of the characters. 
However, previous studies~\cite{Styledubber, speechdubber, HPM, Diff-foley,FoleyCrafter} primarily focus on the generation of either speech or background audio.
To explore this field more comprehensively, we decompose and view the video-to-soundtrack task from two different perspectives, including video-to-audio generation and video-to-speech generation. 

\subsection{Video-to-Speech Generation}

Video-to-speech aims to generate speech, given transcription with both the desired voice specified by reference speech and the desired rhythm presented in the reference video.
 
Previous works extend the conventional text-to-speech (TTS) framework by incorporating video information to improve lip synchronization. For example, VDTTS~\cite{vdtts} introduces pre-trained speaker embeddings to maintain speaker identity consistency, while FaceTTS~\cite{facetts} directly predicts speaker characteristics from video inputs. However, these methods still fall short in lip-sync accuracy. The generated speech often shows misalignment with the lip movements and exhibits low speech quality.
Methods such as HPMDubber~\cite{HPM}, StyleDubber~\cite{Styledubber}, and Speaker2Dubber~\cite{speechdubber} adopt hierarchical or multi-scale prosody modeling to enhance the alignment between speech and video. MCDubber~\cite{mcdubber} further leverages long-range contextual information to improve video-speech synchronization. Nevertheless, these approaches often enforce strict alignment between speech and video, which tends to compromise pronunciation accuracy and speech naturalness, leading to a significant performance gap compared with mainstream TTS models~\cite{styletts2, f5tts, ns2, ns3} in terms of naturalness and speaker similarity.
Alternatively, some methods directly predict speech from video, known as lip-referred dubbing schemes~\cite{V2SFlow, Diffv2s, Lipper}, which typically generate mel-spectrograms from lip sequences via encoder-decoder models. However, due to the high word error rates in these systems, it is challenging to ensure the generation of high-quality speech.
Moreover, current works are mostly limited to constrained scenarios (\textit{e.g.}, lip alignment) and lack the capability for general video-to-speech generation, such as narrative dubbing or voice-over tasks~\cite{Narrativedata}, which require more sophisticated prosodic variations to match the rhythm of the video. 


Therefore, building a universal video-to-speech model that balances speech naturalness, rhythm alignment, and generalization across diverse scenarios remains an under-explored research direction.

\subsection{Video-to-Audio Generation}

The video-to-audio (V2A) task aims to analyze video content and synthesize natural background audio that matches the visual semantics and is consistent in timing.

At present, the V2A methods can be divided into two categories. The first category is transfer learning based on text-to-audio (TTA) generation, including V2A-Mapper~\cite{V2a-mapper}, FoleyCrafter~\cite{FoleyCrafter}, and SonicVisionLM~\cite{Sonicvisionlm}. V2A-Mapper translates visual Contrastive Language–Image Pre-training (CLIP)~\cite{clip} embeddings to Contrastive Language–Audio Pre-training (CLAP)~\cite{clap} space, enabling video-aligned audio generation using a pre-trained TTA generator. FoleyCrafter leverages a pre-trained TTA model to ensure high-quality audio generation and adopts a semantic adapter for semantic alignment and a temporal controller for precise audio-video synchronization. SonicVisionLM first identifies events within the video using vision-language models and then generates audio from captions using a TTA model.
Another category of methods is directly modeling video-to-audio generation from scratch, emphasizing the deep modeling between video and audio. For example, Im2Wav~\cite{im2wav} uses the CLIP feature of the video as a condition and uses an autoregressive method to generate audio tokens frame by frame. Diff-Foley~\cite{Diff-foley} adopts contrastive audio-visual pretraining (CAVP)~\cite{cavp}
to learn more temporally and semantically aligned features and then train a latent diffusion model to achieve high-quality audio synthesis. Frieren~\cite{Frieren} adopts flow-matching structures and a length regulator to model the connection between video and audio explicitly.
While the audio quality and synchronization have been improved a lot, these approaches require a considerable amount of high-quality audio-video data to achieve good generalization across various video scenes.

Although significant progress has been made in video-to-speech or video-to-audio generation, the video-to-soundtrack generation task remains underexplored, with no unified end-to-end framework for simultaneous background audio and speech generation.

\section{DualDub}

\begin{figure*}[htb!]
  \centering
  \includegraphics[width=\linewidth]{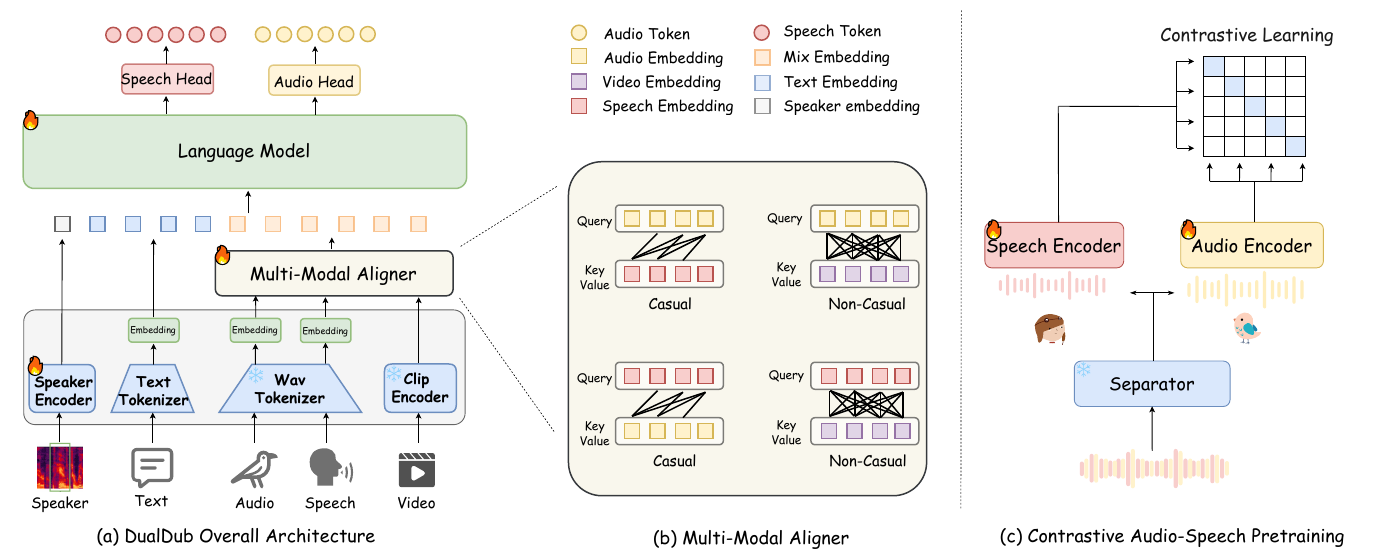}
  \caption{Architecture of the proposed DualDub and Contrastive Audio-Speech Pretraining (CASP). (a) illustrates the overall architecture of DualDub. The gray-shaded area represents the multimodal encoder. The detailed structure of the multimodal aligner is shown in (b). Specifically, the lines indicate which key-value embeddings are visible to the query. The outputs from these four cross-attention modules are summed to generate the final fused representation. (c) depicts the overall architecture of CASP that utilizes Mel-RoFormer to separate the mixed waveform into background audio and speech.
}
  \label{fig_overview}
\end{figure*}

\subsection{Framework Overview}

DualDub aims to generate background audio $\hat{W}^{\text{audio}}$ and dubbing speech $\hat{W}^{\text{speech}}$ harmoniously given a reference speech $R$, a transcript $T$, and a silent video sequence $V$, which can be formulated as
\begin{equation} 
\hat{W}^{\text{speech}}, \hat{W}^{\text{audio}} = \text{DualDub}(R, T, V).
\end{equation}
To achieve this goal, it is essential to handle multimodal input and cross-modal fusion and design a unified framework to achieve the simultaneous generation of audio-speech pairs. As shown in Figure~\ref{fig_overview} (a), DualDub consists of three components. Specifically, the multimodal encoder receives and processes multimodal inputs and converts them into embeddings. The multimodal aligner fuses the audio-speech-video information into a hidden sequence. 
The multimodal language model (LM) then autoregressively processes the speaker prompt, text sequence, and the fused multimodal sequence, with two LM-heads projecting hidden output into two token streams, where one stream is audio tokens, and the other is speech tokens. Finally, a waveform decoder is used to construct high-quality speech and audio waveforms from predicted audio and speech tokens. Thus, speech-audio pairs are generated simultaneously under the guidance of video, text, and speaker prompts, achieving V2ST generation.
Following the above process, each sub-module of DualDub will be introduced sequentially.

\subsection{Multimodal Encoder}

DualDub employs a multimodal encoder architecture for input feature processing, with its core design integrating four key inputs.
For textual input, the system performs tokenization through Byte-Pair Encoding (BPE) to establish semantic representations. These representations provide linguistic content for target speech. The BPE tokenization shortens the length of text representations and improves training efficiency. In addition, it can be expanded to multiple languages, making scaling up easier.
For the input of the reference speech, we randomly intercept 3 seconds of target speech, extract mel-spectrograms as the input of the speaker encoder, and finally get a global speaker embedding through average pooling. The speaker embedding mainly provides the speaker timbre of the target speech.
For audio processing, we adopt a unified WavTokenizer~\cite{wavtokenizer} for tokenizing background audio and speech tracks, obtaining audio and speech representations.
Then, we use two look-up tables to project audio and speech tokens into different embeddings, which isolates the two tracks and reduces model learning complexity.
For video processing, we leverage the visual encoder of the CLIP~\cite{clip} model. The video is regarded as an image sequence input, and each image is sent to the visual encoder to extract image features and then concatenated to form video features. To reduce the amount of computation, one image is taken every three frames in the video stream.

\subsection{Cross-modal Aligner}

In the DualDub framework, the cross-modal aligner is designed to achieve temporal synchronization and cross-modal fusion among video, background audio, and speech, enabling consistency between background audio and speech generation. Inspired by the cross-attention mask strategy~\cite{aligner} for handling intra-modal tasks, the aligner employs a hybrid cross-attention mechanism comprising two distinct interaction patterns to optimize cross-modal synchronization and harmony. 
As shown in Figure~\ref{fig_overview} (b), the hybrid cross-attention mechanism consists of intra-modal causal cross-attention and inter-modal non-causal cross-attention. 

\noindent
\textbf{Intra-modal Causal Cross-attention}. We leverage intra-modal cross-attention to enforce rhythmic coherence within acoustic modalities by perceiving historical embeddings of each other. In addition, we adopt a causal mask to prevent audio and speech streams from seeing each other's future information, matching the pattern during inference. Specifically, we use two causal cross-attention, one using audio embeddings as query and speech embeddings as key-value, while the other uses speech embeddings as query and audio embeddings as key-value to perform cross-attention with the causal mask. The attention results are added to the corresponding speech and audio embeddings to get two outputs. 

\noindent
\textbf{Inter-modal Non-causal Cross-attention}. We design inter-modal non-causal cross-attention to align acoustic modalities with visual modality dynamically. The non-causal mask ensures synchronization of sound with visual actions and enables speech representations to adapt to visual context for context-aware vocal expression. Concretely, we adopt two non-causal cross-attention modules, where audio embeddings and speech embeddings are treated as queries, while video features serve as keys and values. The resulting attention outputs are then integrated with the original queries via residual connections, yielding two vision-aligned acoustic representations.

\subsection{Curriculum Learning}

Taking the representations from the multimodal encoder and aligner, a multimodal language model predicts target speech and audio tokens autoregressively and simultaneously. Hence, the text representations, speaker representations, and aligned representations are concentrated at the time dimension, ensuring the left-to-right modeling of the multimodal language model.

Training multimodal language models directly often demands substantial quantities of paired video-text-speech-audio data to achieve robust performance; however, acquiring such paired datasets is inherently challenging due to their scarcity. Therefore, to address the challenge of data scarcity, we propose a three-stage curriculum learning strategy where the LLM is progressively learning generalization capability through minimal paired data. Each stage employs multi-task training to incrementally develop the model's multimodal generation capabilities, with later phases building upon competencies acquired in earlier ones. The multi-stage curriculum enables systematic knowledge transfer while avoiding catastrophic forgetting.

\noindent
\textbf{Stage 1: Video-to-Audio Foundation}. The model is first adapted to visual inputs using video-audio paired data, where $V$ denotes the visual input and $A = \{ A_1, A_2, \ldots, A_t \}$ represents the corresponding audio tokens. This phase aims to develop the fundamental cross-modal alignment between visual content and audio elements, enabling preliminary video-conditioned audio generation. Formally, the learning objective is to model the conditional probability:

{
\begin{equation}
p(A_t^{\text{audio}}|A_{<t}^{\text{audio}}, V)=\prod_{t=0}^{T}p(A_t^{\text{audio}}|A_{<t}^{\text{audio}}, V;\theta^{LM}, \theta^{Align}),
\end{equation}
}

\noindent
where $A_t^{\text{audio}}$ denotes the audio token at time step $t$, $A_{<t}^{\text{audio}}$ denotes previously generated tokens, $\theta^{LM}$ and $\theta^{Align}$ denote the parameters of the language model and the cross-modal aligner, respectively.
It is worth noting that when performing a video-to-audio task, the text prompt is always set to ``\textit{Generate audio for the video.}'' and the speaker prompt is set to zeros.

\noindent
\textbf{Stage 2: Multimodal Specialization}. We expand DualDub's speech generation capabilities by integrating a text-to-speech (TTS) task while maintaining the V2A task. Specifically, a dual-head architecture is adopted, which is strategically allocated for audio and speech generation tasks. 
During the alternating training of the V2A and TTS tasks, modality-specific masking is applied on model inputs to distinguish task objectives. For the V2A task, the model takes video features $V$ as input, consistent with Stage 1. In contrast, for the TTS task, the video input is replaced with all-zero vectors, and the model conditions on the input text transcription $T$ and reference speech $R$ to synthesize target speech.
This design enables modality-specific modeling while sharing the backbone model parameters, improving parameter efficiency.
Formally, the conditional probability distributions for the two tasks are defined as:
{
\begin{align}
& p(A_t^{\text{audio}}|A_{<t}^{\text{audio}}, V)=\prod_{t=0}^{T}p(A_t^{\text{audio}}|A_{<t}^{\text{audio}}, V;\theta^{LM}, \theta^{Align}),  \\
& p(A_t^{\text{speech}}|A_{<t}^{\text{audio}}, T, R)=\prod_{t=0}^{T}p(A_t^{\text{speech}}|A_{<t}^{\text{speech}}, T, R;\theta^{LM}, \theta^{Align}),
\end{align}
}

\noindent
where $A_t^{\text{audio}}$ denotes the audio token at time step $t$, $A_t^{\text{speech}}$ denotes the speech token at time step $t$. The parameters $\theta^{LM}$ and $\theta^{Align}$ are shared across tasks.

\noindent
\textbf{Stage 3: Video-to-Soundtrack Generation}. In the third stage, the model advances to joint speech and audio generation by learning synchronized audio-speech synthesis through fully paired training data. With audio and speech generation capabilities built by previous stages, Stage 3 just needs a small amount of data to merge these two capabilities and match the inference pattern. During this phase, the multimodal aligner is integrated to achieve synchronization across multiple modalities. Formally, the task added in the third stage can be written as follows:
{
\begin{equation}
\begin{split}
& p(A_t^{\text{speech}}, A_t^{\text{audio}}|A^{\text{speech}}_{<t}, A^{\text{audio}}_{<t}, V, T, R)= \\
&\prod_{t=0}^{T}p(A_t^{\text{speech}}, A_t^{\text{audio}}|A^{\text{speech}}_{<t}, A^{\text{audio}}_{<t}, V, T, R;\theta^{LM}, \theta^{Align}).
\end{split}
\end{equation}
}

\noindent
Notably, we simply leverage the cross-entropy loss as the constraint to optimize token prediction during all stages.

\subsection{Waveform Generation}

The audio codec is mainly designed for audio signal transmission, which primarily pursues the ultimate low bit rate and sacrifices audio quality to a certain extent. To this end, we propose a two-stage decoder to replace the WavTokenizer decoder for better synthesis quality. Motivated by recent advances in flow-matching models~\cite{f5tts, CosyVoice2}, we leverage a variational auto-encoder (VAE) to extract latent representations of the audio modality and then a DiT-based flow-matching network to transform the discrete audio tokens predicted by the language model into the VAE latent space. 

Specifically, we leverage the VAE of StableAudio2~\cite{stableaudio2} to extract acoustic latent representations $Z_{l}$ from raw waveforms. Then, given the token sequence $A$ generated from the language model, we adopt a DiT-based flow matching model $\mathcal{F}(\cdot;\theta)$ to map the tokens into the VAE latent space $Z_{l}$, where $\theta$ denote the learnable parameters.
Formally, we define a continuous-time trajectory $Z_t$ where $t \in [0, 1]$ denotes the time variable. The trajectory starts from the token embedding $Z_0 = \text{Embed}(A)$ and evolves towards the target latent representation $Z_1 = Z_{l}$. The flow-matching model $\mathcal{F}(\cdot;\theta)$ predicts the velocity field $v_{\theta}(Z_t, t)$ to guide the evolution of $Z_t$:
{
\begin{equation} \frac{dZ_t}{dt} = v_{\theta}(Z_t, t) \end{equation}.
}

\noindent
During training, given the target latent $Z_{l}$ extracted from the VAE encoder, we supervise the model by minimizing the flow matching loss:
{
\begin{equation} \mathcal{L}_{FM} = \mathbb{E}_{Z_t, t} \left[ \left| v_{\theta}(Z_t, t) - \frac{Z_{l} - Z_0}{1} \right|_2^2 \right], \end{equation}
where $Z_t$ is obtained by linear interpolation between $Z_0$ and $Z_{l}$: \begin{equation} Z_t = (1-t) \cdot Z_0 + t \cdot Z_{l}. \end{equation}
}
At inference time, we start from the token embedding {
{
\begin{equation} \hat{Z}_{l} = Z_1 = Z_0 + \int_{0}^{1} v_{\theta}(Z_t, t) dt. \end{equation}
}
Finally, the generated $Z_{l}$ is fed into the frozen VAE decoder $\theta_{VAE}$ to synthesize the waveform:
{
\begin{equation} \hat{W}^{wave} = \theta^{VAE}(\hat{Z}_{l}). \end{equation}
}

\noindent
Through this two-stage process, token-to-latent followed by latent-to-wave, we effectively generate high-quality waveform audio from discrete tokens.

\section{DualBench}

\subsection{Benchmark Overview}
To comprehensively and systematically evaluate performance in the video-to-soundtrack task, DualBench assesses results in three different dimensions: generation quality, video synchronization, and audio-speech harmony. In addition, DualBench includes a test dataset for the video-to-soundtrack task.

\subsection{V2ST Test Set}

The absence of dedicated datasets for audio-speech alignment assessment has hindered the development of robust evaluation benchmarks. Our solution leverages the V2C-Animation~\cite{v2c} dataset, initially designed for visual voice conversion, through a rigorous processing pipeline. First, we merge multi-channel audio into a mono-channel format. The Mel-RoFormer~\cite{melroformer} model then separates human speech and background audio, followed by an energy-based filtering stage that removes pairs containing silent speech or audio. Specifically, if either the split audio or speech has an energy lower than -40dB, which is empirically determined, the paired data is discarded.
The V2ST test set is built on the test set of the V2C-Animation dataset~\cite{speechdubber}. Through the above process, we obtain 1,319 curated clips from the original 2,793 clips, with an average duration of 2.65 seconds. 

\subsection{V2ST Metrics}

\noindent
\textbf{Generation Quality}. We access audio and speech quality, respectively. Following mainstream approaches~\cite{audioldm,cosyaudio}, we adopt the audioldm-eval\footnote{\url{https://github.com/haoheliu/audioldm_eval}} tool to compute Fréchet Distance (FD), Fréchet Audio Distance (FAD)~\cite{fad},  Inception Score (IS), and Kullback–Leibler (KL) divergence. FD calculates the distribution distance between the PANN~\cite{panns} feature of the target and generated audio. FAD assesses the similarity of distribution-level VGGish~\cite{vggish} features between generated and reference audio samples. IS evaluates the quality and diversity of generated audio. KL is a reference-dependent metric that measures the divergence between the acoustic event posteriors of ground truth and generated audio. For speech quality evaluation, we adopt speaker similarity~(SIM), word error rate (WER), and UTMOS~\cite{utmos}. SIM assesses the model's capability to clone the voice of an unseen speaker given a reference speech clip. WER measures whether the linguistic content aligns with the provided transcript. UTMOS estimates overall perceptual speech quality. Specifically, we use Whisper-large-v3~\cite{whisper} for WER calculation, WavLM-based speaker verification model~\cite{wavlmsv} for speaker similarity and SpeechMOS\footnote{\url{https://github.com/tarepan/SpeechMOS}} to calculate UTMOS.

\noindent
\textbf{Audio-video Alignment}
We evaluate audio-video alignment with the AV-Align metric~\cite{AValigntool}. Specifically, AV-Align is a metric for evaluating the alignment between audio and video modalities in multimedia data. It assesses synchronization by detecting audio and video peaks and calculating their Intersection over Union (IoU). A higher IoU score indicates better alignment. 

\begin{figure}[t]
  \centering
  \includegraphics[width=0.8\linewidth]{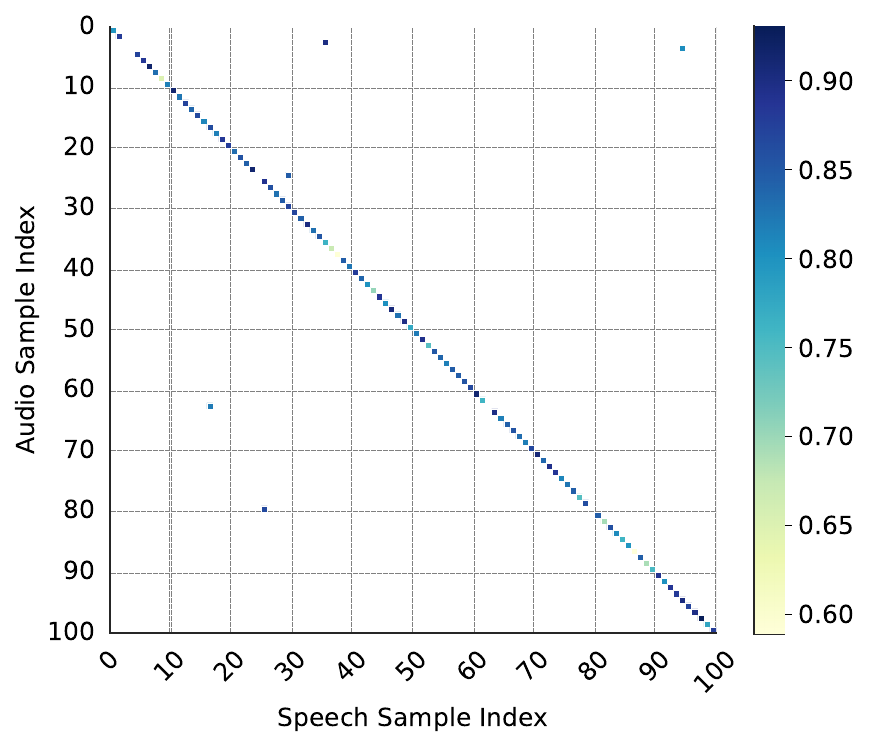}  
  \caption{Heatmap of top-1 retrieval results from contrastive audio-speech pretraining. One hundred samples are randomly selected from the V2ST test set for better visualization.}
  \label{fig_heatmap}
  \vspace{-10pt}
\end{figure}

\noindent
\textbf{Audio-speech Harmony}
As there are no metrics directly evaluating the harmony between audio and speech, we explore various evaluation models and propose DualScore to this end. Specifically, inspired by CLIP~\cite{clip, imagebind}, we introduce contrastive audio-speech pertaining (CASP) to establishing alignment between audio and speech. As shown in Figure~\ref{fig_overview} (c), the architecture of CASP adopts a dual-branch encoder: the audio branch uses the BEATs~\cite{beats} encoder with pre-trained parameters initially in the training phase. In contrast, the parallel speech branch uses the same architecture but is trained from scratch to enable domain-specific adaptation. The two branches converge through an attention-pooling layer to promote cross-modal feature aggregation. During training, the input data is processed by randomly sampling 5-second segments from the audio-speech pairs. After training, the cosine similarity between speech and audio embeddings is defined as DualScore.


To assess the effectiveness of DualScore, we follow CLIP and conduct multi-level retrieval evaluations on the V2ST test set. Specifically, we calculate a Top-$k$ retrieval accuracy (where $k \in \{1, 3, 5\}$), overall 1,319 audio-speech pairs from the V2ST test set. We report the final average accuracy. As shown in the retrieval heatmap Figure~\ref{fig_heatmap}, most correct pairs are concentrated along the diagonal, indicating accurate retrieval performance. Besides, the model achieves 70\% Top-1 accuracy, 90\% Top-3 accuracy, and 95\% Top-5 accuracy, demonstrating the effectiveness of DualScore.


\begin{table}[t]
\centering
\footnotesize
\caption{The corpus used to train models. * means separated and filtered.}
\label{tab_data}
\begin{tabular}{@{}lccc@{}}
\toprule
Dataset & $\sim$Dur. (h) & DualDub & DualBench \\ \midrule
LibriTTS~\cite{librittsdata} & 550 & $\checkmark$ & \\
VGGSound~\cite{vggsounddata} * & 500 & $\checkmark$ & \\
Anim~\cite{Animdataset} * & 100 & $\checkmark$ & $\checkmark$ \\
V2C-Animation\cite{v2c} * & 5 & $\checkmark$ & $\checkmark$ \\
Private multimodal pairs * & 100 & $\checkmark$ & $\checkmark$ \\
Private audio-speech pairs * & 1500 & & $\checkmark$ \\ \bottomrule
\end{tabular}
\end{table}

\begin{table*}[t]
\centering
\setlength{\tabcolsep}{5pt} 
\footnotesize
\caption{Objective evaluation of generated audio on the VGGSound and DualBench test sets. The best and the second best result is shown in \textbf{bold} and by \underline{underlined}.}
\label{table_audio}
\begin{tabular}{lcccccccccc}
\toprule
\multirow{2}{*}{Model} 
& \multicolumn{5}{c}{VGGSound} 
& \multicolumn{5}{c}{DualBench} \\
\cmidrule(lr){2-6} \cmidrule(lr){7-11}
& FD$\downarrow$ & FAD$\downarrow$ & KLD$\downarrow$ & IS$\uparrow$ & AV-Align$\uparrow$ 
& FD$\downarrow$ & FAD$\downarrow$ & KLD$\downarrow$ & IS$\uparrow$ & AV-Align$\uparrow$ \\
\midrule
MMAudio~\cite{mmaudio}      & \textbf{5.60}  & \textbf{1.77} & \textbf{2.10} & \textbf{16.98} & 0.22  & 38.15 & 9.44 & 2.20 & \underline{2.93} & \underline{0.23} \\
FoleyCrafter~\cite{FoleyCrafter} & 26.70 & 2.38 & \underline{2.53} & 9.66  & \textbf{0.25}  & \underline{25.18} & \textbf{4.60} & \underline{1.84} & 2.45 & \underline{0.23} \\
Diff-Foley~\cite{Diff-foley}   & 23.38 & 6.05 & 3.18 & 10.95 & 0.21  & 41.82 & 6.24 & 2.78 & \textbf{5.20} & 0.20 \\
DualDub     & \underline{11.95} & \underline{2.29} & 2.91 & \underline{11.50}  & \underline{0.24}  & \textbf{23.80} & \underline{5.86} & \textbf{1.79} & 2.72 & \textbf{0.25} \\
\bottomrule
\end{tabular}
\end{table*}

\section{Experimental setup}

\subsection{Datasets}
Datasets used in our experiments are as shown in Table~\ref{tab_data}, including three types: V2A data, TTS data, and fully paired data. Specifically, VGGSound~\cite{vggsounddata} is used as the video-to-audio training dataset. LibriTTS~\cite{librittsdata} is used as the text-to-speech training dataset, including three subsets, namely, train-clean-100,train-clean-360, and train-other-500. The multimodal pairs are collected from three primary sources: the Anima dataset~\cite{Animdataset}, the V2C-Animation training set~\cite{v2c}, and our privately processed video data. We also collect the audio-speech pairs from the Internet, which serves to train the evaluation model of DualBench.
Notably, following the data process of DualBench, we separate all audio-speech streams to avoid speech leaks into the background audio.

\subsection{Implement Details}
For the audio-speech tokenization, we utilize the WavTokenizer~\cite{wavtokenizer}, a codec model, for unified tokenization, operating at a 24~kHz sampling rate with a 600 hop size and achieving a code rate of 40 tokens per second\footnote{\url{https://huggingface.co/novateur/WavTokenizer-large-unify-40token}}.
For video feature extraction, we employ a CLIP-based~\cite{clip} visual encoder with the ViT-B/32 configuration to process video inputs.
We apply nearest-neighbor interpolation to resample the video features to match the frame rate of the audio tokens.
For text tokenization, we adopt tiktoken BPE\footnote{\url{https://github.com/openai/tiktoken}} for text input. 
The DualDub model is implemented using the LLaMA~\cite{llama} architecture with 24 layers, 16 attention heads, and a hidden dimension of 2048.  
The waveform decoder adopts the pre-trained VAE from StableAudio2~\cite{stableaudio2}, and the DiT-based flow-matching model consists of 8 transformer layers, 16 attention heads, and a hidden dimension of 2048.
Considering the different frame rates of input tokens and latent, we use the nearest neighbor interpolation method to interpolate the token embedding to maintain the same frame rate with latent representations.

All training phases are conducted on eight NVIDIA A800 GPUs using a cosine learning rate scheduler with 4,000 warm-up steps. We use the Adam optimizer with $\beta_1 = 0.9$, $\beta_2 = 0.98$, and $\epsilon = 10^{-9}$. The model undergoes three training stages with distinct learning rate ranges: the first stage employs rates between $2e^{-6}$ and $2e^{-4}$, the second stage uses $2e^{-7}$ to $2e^{-4}$, and the final stage is performed with learning rates from $2e^{-7}$ to $2e^{-5}$. Each training stage maintains a global batch size of 192 and completes 40 epochs of training. The multimodal aligner and language model are optimized, while the CLIP and WavTokenizer are kept frozen during all training phases.

\subsection{Comparison Systems}
As this work is the first attempt to achieve V2ST generation, there are no existing methods directly comparable to DualDub. Therefore, we first choose representative V2A and V2S models and leverage their open-source models for single-element generation comparison. Then, we construct baselines by concatenating the best V2A and V2S models to compare dual-element generation.

\begin{itemize}
    \item \textbf{MMAudio}~\cite{mmaudio}: MMAudio is built on the MM-DiT block design from SD3~\cite{stablediffusion3} and utilizes both video-to-audio and text-to-audio datasets. We use the MMAudio-large for comparison.
    \item \textbf{Diff-Foley}~\cite{Diff-foley}: Diff-Foley adopts an LDM with CAVP-aligned visual features on spectrogram latent space.
    \item \textbf{FoleyCrafter}~\cite{FoleyCrafter}: FoleyCrafter integrates a learnable module into a pre-trained text-to-audio model. During inference, we only use video as input without text.
    \item \textbf{Speaker2Dubber}~\cite{speechdubber}: Speaker2Dubber is a two-stage training framework where prosody and duration consistency are modeled especially.
    \item \textbf{StyleDubber}~\cite{Styledubber}: StyleDubber combines phoneme-level and utterance-level modeling for prosody and timbre representation.
    \item \textbf{HPMDubber}~\cite{HPM}: HPMDubber improves contextual prosody through facial feature modeling and aggregation of contexts between sentences.
\end{itemize}

\section{Experimental Results}

\subsection{V2A and V2S Generation}

\noindent
\textbf{Video-to-Audio.} Notably, V2A generation is widely tested on the VGGSound test set. Therefore, we evaluate both DualBench and the VGGSound test sets, as shown in Table~\ref{table_audio}. In the in-domain test set (VGGSound), DualDub achieves satisfactory performance on both audio quality and audio-video alignment metrics. MMAudio achieves the best audio quality, significantly outperforming FoleyCrafter and Diff-Foley, while FoleyCrafter achieves better performance in audio-video alignment. Moreover, DualDub obtains a relatively low IS score, possibly due to the limitation of the WavTokenizer. Although we adopt a two-stage decoder to improve audio quality, it is still inferior to fully end-to-end models.
In the out-of-domain test set (DualBench), DualDub achieves better performance, slightly surpassing FoleyCrafter. Interestingly, compared to the results of the in-domain test set, all models exhibit some performance degradation. However, FoleyCrafter maintains relatively stable performance, likely benefiting from the pre-trained T2A model with better generalization ability. Overall, DualDub achieves competitive performance compared to these SOTA V2A models, demonstrating the powerful audio generation capability even though DualDub is trained mainly for the V2ST task.

\noindent
\textbf{Video-to-Speech.} Table~\ref{table_obj_speech_imbench} compares our V2S results on the DualBench. DualDub achieves superior performance, significantly outperforming other V2S models in terms of WER, SIM, and UTMOS. This result indicates that DualDub synthesizes intelligible and natural speech, aligning with the advantage of leveraging language models for zero-shot text-to-speech tasks and demonstrating the strong speech generation ability of DualDub.

\begin{table}[t]
\centering  
\footnotesize
\caption{Objective speech quality metrics on the DualBench test set.}
\label{table_obj_speech_imbench}
\begin{tabular}{lccc}  
\toprule
Model        & WER(\%)↓ & SIM↑ & UTMOS↑ \\ \midrule
GroundTruth  & 35.62    & 1.00 & 2.02   \\\midrule
HPMDubber~\cite{HPM}          & 20.96    & 0.74 & 1.31   \\
Speaker2Dubber~\cite{speechdubber}  & 24.92    & 0.82 & 2.28   \\
StyleDubber~\cite{Styledubber}      & 30.57    & 0.82 & 1.97   \\
DualDub     & \textbf{12.74}    & \textbf{0.84} & \textbf{2.70}   \\ \bottomrule
\end{tabular}
\end{table}

\subsection{V2ST Generation}

We conduct objective and subjective experiments to evaluate each baseline. Specifically, apart from the objective metric, DualScore, we invited 10 non-expert listeners and 10 with professional audio knowledge for subjective evaluation. We require listeners to evaluate audio samples in three aspects: (1) overall audio quality (OAQ), which evaluates the rationality and quality of the audio; (2) overall speech quality (OSQ), which evaluates the rationality and quality of the speech; and (3) audio-speech correspondence and harmony (ASCH), which evaluates the semantic and rhythmic consistency between the audio and speech.
We use 16 sample pairs to evaluate the metrics. All these subjective metrics are based on a 5-point scale and reported with the 95\% confidence intervals.

\begin{table*}[t]
\caption{Results of ablation study on DualBench.}
\label{tableablation}
\centering
\footnotesize
\begin{tabular}{@{}lccccccccc@{}}
\toprule
\multirow{2}{*}{Model}                    & \multicolumn{4}{c}{Audio Quality} & Audio-Video alignment & \multicolumn{3}{c}{Speech Quality} & Harmonization assessment \\ \cmidrule(l){2-5} \cmidrule(l){6-6} \cmidrule(l){7-9}  \cmidrule(l){10-10}
                                          & FD↓  & FAD↓ & KLD↓ & IS↑  & AV-Align↑             & SSIM↑    & WER(\%)↓    & UTMOS↑   & DualScore ↑                \\ \midrule
DualDub                                 & \textbf{23.08}    & \textbf{5.86}     & \textbf{1.79} & \textbf{2.72} & \textbf{0.25}                  & 0.84     & 12.74       & \textbf{2.70}     & \textbf{0.59}                        \\
\quad w/o curriculum learning             & 31.61    & 9.61    & 2.60 & 2.33 & 0.21                  & 0.79     & 15.40       & 2.58     & 0.56                        \\
\quad w/o cross-modal aligner & 24.26    & 6.78     & 1.87 & 2.69 & 0.18                  & \textbf{0.85}        & \textbf{11.96}           & 2.69        & 0.36                        \\
\quad w/o flow-matching decoder                    & 42.61    & 13.61    & 6.74 & 1.90 & 0.19                  & 0.82     & 13.55       & 2.62        & 0.41                        \\ \bottomrule
\end{tabular}
\end{table*}

\begin{table}[t]
\caption{Objective and subjective metrics for soundtrack generation on the DualBench test set. S2D denotes Speaker2Dubber, SD denotes StyleDubber, MM denotes MMAudio, and FC denotes FoleyCrafter.}
\label{table_soundtrackgeneration}
\footnotesize
\begin{tabular}{@{}lcccc@{}}
\toprule
\multirow{2}{*}{Model}        & \multicolumn{1}{c}{Objective} & \multicolumn{3}{c}{Subjective}          \\ 
\cmidrule(lr){2-2} \cmidrule(lr){3-5}
                              & DualScore↑ & OAQ↑ & OSQ↑ & ASCH↑ \\ 
\midrule
GroundTruth                   & 0.84 & 2.69 ± 0.08 & 3.50 ± 0.09 & 3.50 ± 0.10 \\ \midrule 
S2D~\cite{speechdubber} + MM~\cite{mmaudio}      & 0.19 & 1.94 ± 0.08 & 2.62 ± 0.11 & 1.44 ± 0.06 \\
SD~\cite{Styledubber} + MM~\cite{mmaudio}        & 0.23 & 1.94 ± 0.08 & 2.81 ± 0.13 & 1.50 ± 0.06 \\
S2D~\cite{speechdubber} + FC~\cite{FoleyCrafter} & 0.23 & 2.12 ± 0.07 & 2.62 ± 0.11 & 1.56 ± 0.05 \\
SD~\cite{Styledubber} + FC~\cite{FoleyCrafter}   & 0.30 & 2.12 ± 0.07 & 2.81 ± 0.13 & 1.69 ± 0.06 \\
DualDub                       & \textbf{0.59} & \textbf{2.75 ± 0.08} & \textbf{3.06 ± 0.07} & \textbf{3.44 ± 0.10} \\
\bottomrule
\end{tabular}
\end{table}

\noindent
\textbf{Objective Evaluation.} As shown in Table~\ref{table_soundtrackgeneration}, DualDub achieves state-of-the-art results on the DualBench benchmark. In contrast, the scores obtained by simply concatenating the separately generated background audio and speech are significantly lower than those of DualDub. These independent V2A and V2S models can not perceive each other during generation, often leading to conflicts in rhythm. However, DualDub generates background audio and speech simultaneously in a unified model, which greatly reduces the risk of conflicts. 


\noindent
\textbf{Subjective Evaluation.} As shown in Table~\ref{table_soundtrackgeneration}, in terms of subjective evaluation on OAQ, OSQ, and ASCH, DualDub significantly outperforms other baseline models based on concatenation. Specifically, the background audio samples generated by the V2A baselines often contain meaningless vocal sounds, which leads to severe disharmony when combined with the output of V2S models. Moreover, the background audio in these baselines tends to overwhelm the speech with its volume, further degrading speech intelligibility and leading to low subjective perception scores.

\subsection{Ablation Study}

We conduct an ablation study to assess the effectiveness of the training strategy and each component by evaluating their impact on soundtrack generation. Results are shown in Table~\ref {tableablation}.

\noindent
\textbf{Curriculum Learning}. Skipping stage 1 and stage 2 pretraining and directly performing end-to-end optimization leads to significant performance degradation in speech quality, which indicates that single-stage training can not optimize audio and speech well. Moreover, we observe that single-stage training encounters unstable and slow convergence, while curriculum learning exhibits stable optimization and improved generation quality.


\noindent
\textbf{Cross-modal Aligner}. Removing the cross-modal aligner slightly impacts generation quality. However, without perception of other modalities, the synthetic audio and speech get the lowest AV-Align score and DualScore, showing that the cross-modal aligner enhances the synchronization among modalities.


\noindent
\textbf{Flow-matching Decoder}. Replacing the flow-matching decoder with the original WavTokenizer's decoder significantly decreases the audio quality and even introduces some background noise. Meanwhile, speech quality only drops slightly, potentially due to WavTokenizer's better capability of speech generation. However, the significant drop in audio quality leads to a lower DualScore. These results demonstrate that the flow-matching decoder plays a crucial role in DualDub for achieving high-quality audio generation.


\section{Conclusion}
This paper proposes DualDub, a framework that generates highly synchronized and harmonious soundtracks for videos.
Specifically, we propose a cross-modal aligner incorporating non-causal and causal cross-attention to ensure multimodal synchronization and harmony. Furthermore, we introduce curriculum learning that effectively leverages minimal multimodal data to extend the capabilities progressively. 
We also establish an open-source benchmark, DualBench, for systematically evaluating the video-to-soundtrack task. 
Extensive experiments demonstrate that DualDub achieves state-of-the-art performance in generating coherent and harmonious soundtracks, marking a promising approach for V2ST generation.

\section{Limitations}
Although DualDub achieves better performance than comparison models in the V2ST task, there are some limitations. The first is the tokens extracted by WavTokenizer, which lack fine-grained acoustic details, hindering the quality and naturalness of generated speech and audio. Recent advances in language modeling ~\cite{kalle, melle} leverage continuous representations and demonstrate superior generation capability, suggesting a promising alternative to overcome the limitations of discrete tokenization. The second is Mel-RoFormer, which sometimes separates empty audio or speech segments, resulting in a large amount of data being filtered out. Therefore, an improved separation model could enhance overall performance.  

\clearpage

\bibliographystyle{ACM-Reference-Format}
\bibliography{sample-base}

\begin{thebibliography}{58}


\ifx \showCODEN    \undefined \def \showCODEN     #1{\unskip}     \fi
\ifx \showDOI      \undefined \def \showDOI       #1{#1}\fi
\ifx \showISBNx    \undefined \def \showISBNx     #1{\unskip}     \fi
\ifx \showISBNxiii \undefined \def \showISBNxiii  #1{\unskip}     \fi
\ifx \showISSN     \undefined \def \showISSN      #1{\unskip}     \fi
\ifx \showLCCN     \undefined \def \showLCCN      #1{\unskip}     \fi
\ifx \shownote     \undefined \def \shownote      #1{#1}          \fi
\ifx \showarticletitle \undefined \def \showarticletitle #1{#1}   \fi
\ifx \showURL      \undefined \def \showURL       {\relax}        \fi
\providecommand\bibfield[2]{#2}
\providecommand\bibinfo[2]{#2}
\providecommand\natexlab[1]{#1}
\providecommand\showeprint[2][]{arXiv:#2}

\bibitem[Cai et~al\mbox{.}(2024)]%
        {Animdataset}
\bibfield{author}{\bibinfo{person}{Kevin Cai}, \bibinfo{person}{Chonghua Liu}, {and} \bibinfo{person}{David~M Chan}.} \bibinfo{year}{2024}\natexlab{}.
\newblock \showarticletitle{Anim-400K: A Large-Scale Dataset for Automated End to End Dubbing of Video}. In \bibinfo{booktitle}{\emph{ICASSP 2024-2024 IEEE International Conference on Acoustics, Speech and Signal Processing (ICASSP)}}. IEEE, \bibinfo{pages}{11796--11800}.
\newblock


\bibitem[Chen et~al\mbox{.}(2020)]%
        {vggsounddata}
\bibfield{author}{\bibinfo{person}{Honglie Chen}, \bibinfo{person}{Weidi Xie}, \bibinfo{person}{Andrea Vedaldi}, {and} \bibinfo{person}{Andrew Zisserman}.} \bibinfo{year}{2020}\natexlab{}.
\newblock \showarticletitle{Vggsound: A large-scale audio-visual dataset}. In \bibinfo{booktitle}{\emph{ICASSP 2020-2020 IEEE International Conference on Acoustics, Speech and Signal Processing (ICASSP)}}. IEEE, \bibinfo{pages}{721--725}.
\newblock


\bibitem[Chen et~al\mbox{.}(2022b)]%
        {v2c}
\bibfield{author}{\bibinfo{person}{Qi Chen}, \bibinfo{person}{Mingkui Tan}, \bibinfo{person}{Yuankai Qi}, \bibinfo{person}{Jiaqiu Zhou}, \bibinfo{person}{Yuanqing Li}, {and} \bibinfo{person}{Qi Wu}.} \bibinfo{year}{2022}\natexlab{b}.
\newblock \showarticletitle{V2C: Visual voice cloning}. In \bibinfo{booktitle}{\emph{Proceedings of the IEEE/CVF Conference on Computer Vision and Pattern Recognition}}. \bibinfo{pages}{21242--21251}.
\newblock


\bibitem[Chen et~al\mbox{.}(2022c)]%
        {beats}
\bibfield{author}{\bibinfo{person}{Sanyuan Chen}, \bibinfo{person}{Yu Wu}, \bibinfo{person}{Chengyi Wang}, \bibinfo{person}{Shujie Liu}, \bibinfo{person}{Daniel Tompkins}, \bibinfo{person}{Zhuo Chen}, {and} \bibinfo{person}{Furu Wei}.} \bibinfo{year}{2022}\natexlab{c}.
\newblock \showarticletitle{Beats: Audio pre-training with acoustic tokenizers}.
\newblock \bibinfo{journal}{\emph{arXiv preprint arXiv:2212.09058}} (\bibinfo{year}{2022}).
\newblock


\bibitem[Chen et~al\mbox{.}(2024)]%
        {f5tts}
\bibfield{author}{\bibinfo{person}{Yushen Chen}, \bibinfo{person}{Zhikang Niu}, \bibinfo{person}{Ziyang Ma}, \bibinfo{person}{Keqi Deng}, \bibinfo{person}{Chunhui Wang}, \bibinfo{person}{Jian Zhao}, \bibinfo{person}{Kai Yu}, {and} \bibinfo{person}{Xie Chen}.} \bibinfo{year}{2024}\natexlab{}.
\newblock \showarticletitle{{F5-TTS:} {A} Fairytaler that Fakes Fluent and Faithful Speech with Flow Matching}.
\newblock \bibinfo{journal}{\emph{CoRR}}  \bibinfo{volume}{abs/2410.06885} (\bibinfo{year}{2024}).
\newblock


\bibitem[Chen et~al\mbox{.}(2022a)]%
        {wavlmsv}
\bibfield{author}{\bibinfo{person}{Zhengyang Chen}, \bibinfo{person}{Sanyuan Chen}, \bibinfo{person}{Yu Wu}, \bibinfo{person}{Yao Qian}, \bibinfo{person}{Chengyi Wang}, \bibinfo{person}{Shujie Liu}, \bibinfo{person}{Yanmin Qian}, {and} \bibinfo{person}{Michael Zeng}.} \bibinfo{year}{2022}\natexlab{a}.
\newblock \showarticletitle{Large-scale self-supervised speech representation learning for automatic speaker verification}. In \bibinfo{booktitle}{\emph{ICASSP 2022-2022 IEEE International Conference on Acoustics, Speech and Signal Processing (ICASSP)}}. IEEE, \bibinfo{pages}{6147--6151}.
\newblock


\bibitem[Cheng et~al\mbox{.}(2024)]%
        {mmaudio}
\bibfield{author}{\bibinfo{person}{Ho~Kei Cheng}, \bibinfo{person}{Masato Ishii}, \bibinfo{person}{Akio Hayakawa}, \bibinfo{person}{Takashi Shibuya}, \bibinfo{person}{Alexander~G. Schwing}, {and} \bibinfo{person}{Yuki Mitsufuji}.} \bibinfo{year}{2024}\natexlab{}.
\newblock \showarticletitle{Taming Multimodal Joint Training for High-Quality Video-to-Audio Synthesis}.
\newblock \bibinfo{journal}{\emph{CoRR}}  \bibinfo{volume}{abs/2412.15322} (\bibinfo{year}{2024}).
\newblock


\bibitem[Choi et~al\mbox{.}(2023)]%
        {Diffv2s}
\bibfield{author}{\bibinfo{person}{Jeongsoo Choi}, \bibinfo{person}{Joanna Hong}, {and} \bibinfo{person}{Yong~Man Ro}.} \bibinfo{year}{2023}\natexlab{}.
\newblock \showarticletitle{Diffv2s: Diffusion-based video-to-speech synthesis with vision-guided speaker embedding}. In \bibinfo{booktitle}{\emph{Proceedings of the IEEE/CVF International Conference on Computer Vision}}. \bibinfo{pages}{7812--7821}.
\newblock


\bibitem[Choi et~al\mbox{.}(2024)]%
        {V2SFlow}
\bibfield{author}{\bibinfo{person}{Jeongsoo Choi}, \bibinfo{person}{Ji{-}Hoon Kim}, \bibinfo{person}{Jinyu Li}, \bibinfo{person}{Joon~Son Chung}, {and} \bibinfo{person}{Shujie Liu}.} \bibinfo{year}{2024}\natexlab{}.
\newblock \showarticletitle{V2SFlow: Video-to-Speech Generation with Speech Decomposition and Rectified Flow}.
\newblock \bibinfo{journal}{\emph{CoRR}}  \bibinfo{volume}{abs/2411.19486} (\bibinfo{year}{2024}).
\newblock


\bibitem[Cong et~al\mbox{.}(2023)]%
        {HPM}
\bibfield{author}{\bibinfo{person}{Gaoxiang Cong}, \bibinfo{person}{Liang Li}, \bibinfo{person}{Yuankai Qi}, \bibinfo{person}{Zheng-Jun Zha}, \bibinfo{person}{Qi Wu}, \bibinfo{person}{Wenyu Wang}, \bibinfo{person}{Bin Jiang}, \bibinfo{person}{Ming-Hsuan Yang}, {and} \bibinfo{person}{Qingming Huang}.} \bibinfo{year}{2023}\natexlab{}.
\newblock \showarticletitle{Learning to dub movies via hierarchical prosody models}. In \bibinfo{booktitle}{\emph{Proceedings of the IEEE/CVF Conference on Computer Vision and Pattern Recognition}}. \bibinfo{pages}{14687--14697}.
\newblock


\bibitem[Cong et~al\mbox{.}(2024)]%
        {Styledubber}
\bibfield{author}{\bibinfo{person}{Gaoxiang Cong}, \bibinfo{person}{Yuankai Qi}, \bibinfo{person}{Liang Li}, \bibinfo{person}{Amin Beheshti}, \bibinfo{person}{Zhedong Zhang}, \bibinfo{person}{Anton van~den Hengel}, \bibinfo{person}{Ming{-}Hsuan Yang}, \bibinfo{person}{Chenggang Yan}, {and} \bibinfo{person}{Qingming Huang}.} \bibinfo{year}{2024}\natexlab{}.
\newblock \showarticletitle{StyleDubber: Towards Multi-Scale Style Learning for Movie Dubbing}. In \bibinfo{booktitle}{\emph{Findings of the Association for Computational Linguistics, {ACL} 2024, Bangkok, Thailand and virtual meeting, August 11-16, 2024}}, \bibfield{editor}{\bibinfo{person}{Lun{-}Wei Ku}, \bibinfo{person}{Andre Martins}, {and} \bibinfo{person}{Vivek Srikumar}} (Eds.). \bibinfo{publisher}{Association for Computational Linguistics}, \bibinfo{pages}{6767--6779}.
\newblock


\bibitem[Du et~al\mbox{.}(2023)]%
        {lm4}
\bibfield{author}{\bibinfo{person}{Yuexi Du}, \bibinfo{person}{Ziyang Chen}, \bibinfo{person}{Justin Salamon}, \bibinfo{person}{Bryan~C. Russell}, {and} \bibinfo{person}{Andrew Owens}.} \bibinfo{year}{2023}\natexlab{}.
\newblock \showarticletitle{Conditional Generation of Audio from Video via Foley Analogies}. In \bibinfo{booktitle}{\emph{{IEEE/CVF} Conference on Computer Vision and Pattern Recognition, {CVPR} 2023, Vancouver, BC, Canada, June 17-24, 2023}}. \bibinfo{publisher}{{IEEE}}, \bibinfo{pages}{2426--2436}.
\newblock


\bibitem[Du et~al\mbox{.}(2024)]%
        {CosyVoice2}
\bibfield{author}{\bibinfo{person}{Zhihao Du}, \bibinfo{person}{Yuxuan Wang}, \bibinfo{person}{Qian Chen}, \bibinfo{person}{Xian Shi}, \bibinfo{person}{Xiang Lv}, \bibinfo{person}{Tianyu Zhao}, \bibinfo{person}{Zhifu Gao}, \bibinfo{person}{Yexin Yang}, \bibinfo{person}{Changfeng Gao}, \bibinfo{person}{Hui Wang}, \bibinfo{person}{Fan Yu}, \bibinfo{person}{Huadai Liu}, \bibinfo{person}{Zhengyan Sheng}, \bibinfo{person}{Yue Gu}, \bibinfo{person}{Chong Deng}, \bibinfo{person}{Wen Wang}, \bibinfo{person}{Shiliang Zhang}, \bibinfo{person}{Zhijie Yan}, {and} \bibinfo{person}{Jingren Zhou}.} \bibinfo{year}{2024}\natexlab{}.
\newblock \showarticletitle{CosyVoice 2: Scalable Streaming Speech Synthesis with Large Language Models}.
\newblock \bibinfo{journal}{\emph{CoRR}}  \bibinfo{volume}{abs/2412.10117} (\bibinfo{year}{2024}).
\newblock


\bibitem[Elizalde et~al\mbox{.}(2023)]%
        {clap}
\bibfield{author}{\bibinfo{person}{Benjamin Elizalde}, \bibinfo{person}{Soham Deshmukh}, \bibinfo{person}{Mahmoud~Al Ismail}, {and} \bibinfo{person}{Huaming Wang}.} \bibinfo{year}{2023}\natexlab{}.
\newblock \showarticletitle{{CLAP} Learning Audio Concepts from Natural Language Supervision}. In \bibinfo{booktitle}{\emph{{IEEE} International Conference on Acoustics, Speech and Signal Processing {ICASSP} 2023, Rhodes Island, Greece, June 4-10, 2023}}. \bibinfo{publisher}{{IEEE}}, \bibinfo{pages}{1--5}.
\newblock


\bibitem[Esser et~al\mbox{.}({[n.\,d.]})]%
        {stablediffusion3}
\bibfield{author}{\bibinfo{person}{Patrick Esser}, \bibinfo{person}{Sumith Kulal}, \bibinfo{person}{Andreas Blattmann}, \bibinfo{person}{Rahim Entezari}, \bibinfo{person}{Jonas M{\"u}ller}, \bibinfo{person}{Harry Saini}, \bibinfo{person}{Yam Levi}, \bibinfo{person}{Dominik Lorenz}, \bibinfo{person}{Axel Sauer}, \bibinfo{person}{Frederic Boesel}, {et~al\mbox{.}}} \bibinfo{year}{[n.\,d.]}\natexlab{}.
\newblock \showarticletitle{Scaling rectified flow transformers for high-resolution image synthesis}. In \bibinfo{booktitle}{\emph{Forty-first international conference on machine learning}}.
\newblock


\bibitem[Evans et~al\mbox{.}(2024)]%
        {stableaudio2}
\bibfield{author}{\bibinfo{person}{Zach Evans}, \bibinfo{person}{CJ Carr}, \bibinfo{person}{Josiah Taylor}, \bibinfo{person}{Scott~H. Hawley}, {and} \bibinfo{person}{Jordi Pons}.} \bibinfo{year}{2024}\natexlab{}.
\newblock \showarticletitle{Fast Timing-Conditioned Latent Audio Diffusion}. In \bibinfo{booktitle}{\emph{Forty-first International Conference on Machine Learning, {ICML} 2024, Vienna, Austria, July 21-27, 2024}}. \bibinfo{publisher}{OpenReview.net}.
\newblock


\bibitem[Fu et~al\mbox{.}(2024)]%
        {Narrativedata}
\bibfield{author}{\bibinfo{person}{Ruibo Fu}, \bibinfo{person}{Shuchen Shi}, \bibinfo{person}{Hongming Guo}, \bibinfo{person}{Tao Wang}, \bibinfo{person}{Chunyu Qiang}, \bibinfo{person}{Zhengqi Wen}, \bibinfo{person}{Jianhua Tao}, \bibinfo{person}{Xin Qi}, \bibinfo{person}{Yi Lu}, \bibinfo{person}{Xiaopeng Wang}, \bibinfo{person}{Zhiyong Wang}, \bibinfo{person}{Yukun Liu}, \bibinfo{person}{Xuefei Liu}, \bibinfo{person}{Shuai Zhang}, {and} \bibinfo{person}{Guanjun Li}.} \bibinfo{year}{2024}\natexlab{}.
\newblock \showarticletitle{{MINT:} a Multi-modal Image and Narrative Text Dubbing Dataset for Foley Audio Content Planning and Generation}.
\newblock \bibinfo{journal}{\emph{CoRR}}  \bibinfo{volume}{abs/2406.10591} (\bibinfo{year}{2024}).
\newblock


\bibitem[Girdhar et~al\mbox{.}(2023)]%
        {imagebind}
\bibfield{author}{\bibinfo{person}{Rohit Girdhar}, \bibinfo{person}{Alaaeldin El{-}Nouby}, \bibinfo{person}{Zhuang Liu}, \bibinfo{person}{Mannat Singh}, \bibinfo{person}{Kalyan~Vasudev Alwala}, \bibinfo{person}{Armand Joulin}, {and} \bibinfo{person}{Ishan Misra}.} \bibinfo{year}{2023}\natexlab{}.
\newblock \showarticletitle{ImageBind One Embedding Space to Bind Them All}. In \bibinfo{booktitle}{\emph{{IEEE/CVF} Conference on Computer Vision and Pattern Recognition, {CVPR} 2023, Vancouver, BC, Canada, June 17-24, 2023}}. \bibinfo{publisher}{{IEEE}}, \bibinfo{pages}{15180--15190}.
\newblock


\bibitem[Gong et~al\mbox{.}(2022)]%
        {cavp}
\bibfield{author}{\bibinfo{person}{Yuan Gong}, \bibinfo{person}{Andrew Rouditchenko}, \bibinfo{person}{Alexander~H Liu}, \bibinfo{person}{David Harwath}, \bibinfo{person}{Leonid Karlinsky}, \bibinfo{person}{Hilde Kuehne}, {and} \bibinfo{person}{James Glass}.} \bibinfo{year}{2022}\natexlab{}.
\newblock \showarticletitle{Contrastive audio-visual masked autoencoder}.
\newblock \bibinfo{journal}{\emph{arXiv preprint arXiv:2210.07839}} (\bibinfo{year}{2022}).
\newblock


\bibitem[Hassid et~al\mbox{.}(2022)]%
        {vdtts}
\bibfield{author}{\bibinfo{person}{Michael Hassid}, \bibinfo{person}{Michelle~Tadmor Ramanovich}, \bibinfo{person}{Brendan Shillingford}, \bibinfo{person}{Miaosen Wang}, \bibinfo{person}{Ye Jia}, {and} \bibinfo{person}{Tal Remez}.} \bibinfo{year}{2022}\natexlab{}.
\newblock \showarticletitle{More than Words: In-the-Wild Visually-Driven Prosody for Text-to-Speech}. In \bibinfo{booktitle}{\emph{{IEEE/CVF} Conference on Computer Vision and Pattern Recognition, {CVPR} 2022, New Orleans, LA, USA, June 18-24, 2022}}. \bibinfo{publisher}{{IEEE}}, \bibinfo{pages}{10577--10587}.
\newblock


\bibitem[Hershey et~al\mbox{.}(2017)]%
        {vggish}
\bibfield{author}{\bibinfo{person}{Shawn Hershey}, \bibinfo{person}{Sourish Chaudhuri}, \bibinfo{person}{Daniel~PW Ellis}, \bibinfo{person}{Jort~F Gemmeke}, \bibinfo{person}{Aren Jansen}, \bibinfo{person}{R~Channing Moore}, \bibinfo{person}{Manoj Plakal}, \bibinfo{person}{Devin Platt}, \bibinfo{person}{Rif~A Saurous}, \bibinfo{person}{Bryan Seybold}, {et~al\mbox{.}}} \bibinfo{year}{2017}\natexlab{}.
\newblock \showarticletitle{CNN architectures for large-scale audio classification}. In \bibinfo{booktitle}{\emph{2017 ieee international conference on acoustics, speech and signal processing (icassp)}}. IEEE, \bibinfo{pages}{131--135}.
\newblock


\bibitem[Huang et~al\mbox{.}(2024)]%
        {audiogpt}
\bibfield{author}{\bibinfo{person}{Rongjie Huang}, \bibinfo{person}{Mingze Li}, \bibinfo{person}{Dongchao Yang}, \bibinfo{person}{Jiatong Shi}, \bibinfo{person}{Xuankai Chang}, \bibinfo{person}{Zhenhui Ye}, \bibinfo{person}{Yuning Wu}, \bibinfo{person}{Zhiqing Hong}, \bibinfo{person}{Jiawei Huang}, \bibinfo{person}{Jinglin Liu}, \bibinfo{person}{Yi Ren}, \bibinfo{person}{Yuexian Zou}, \bibinfo{person}{Zhou Zhao}, {and} \bibinfo{person}{Shinji Watanabe}.} \bibinfo{year}{2024}\natexlab{}.
\newblock \showarticletitle{AudioGPT: Understanding and Generating Speech, Music, Sound, and Talking Head}. In \bibinfo{booktitle}{\emph{Thirty-Eighth {AAAI} Conference on Artificial Intelligence, {AAAI} 2024, Thirty-Sixth Conference on Innovative Applications of Artificial Intelligence, {IAAI} 2024, Fourteenth Symposium on Educational Advances in Artificial Intelligence, {EAAI} 2014, February 20-27, 2024, Vancouver, Canada}}, \bibfield{editor}{\bibinfo{person}{Michael~J. Wooldridge}, \bibinfo{person}{Jennifer~G. Dy}, {and} \bibinfo{person}{Sriraam Natarajan}} (Eds.). \bibinfo{publisher}{{AAAI} Press}, \bibinfo{pages}{23802--23804}.
\newblock


\bibitem[Iashin and Rahtu(2021)]%
        {iashin2021taming-specvqgan}
\bibfield{author}{\bibinfo{person}{Vladimir Iashin} {and} \bibinfo{person}{Esa Rahtu}.} \bibinfo{year}{2021}\natexlab{}.
\newblock \showarticletitle{Taming Visually Guided Sound Generation}. In \bibinfo{booktitle}{\emph{British Machine Vision Conference}}.
\newblock


\bibitem[Ji et~al\mbox{.}(2024)]%
        {wavtokenizer}
\bibfield{author}{\bibinfo{person}{Shengpeng Ji}, \bibinfo{person}{Ziyue Jiang}, \bibinfo{person}{Xize Cheng}, \bibinfo{person}{Yifu Chen}, \bibinfo{person}{Minghui Fang}, \bibinfo{person}{Jialong Zuo}, \bibinfo{person}{Qian Yang}, \bibinfo{person}{Ruiqi Li}, \bibinfo{person}{Ziang Zhang}, \bibinfo{person}{Xiaoda Yang}, \bibinfo{person}{Rongjie Huang}, \bibinfo{person}{Yidi Jiang}, \bibinfo{person}{Qian Chen}, \bibinfo{person}{Siqi Zheng}, \bibinfo{person}{Wen Wang}, {and} \bibinfo{person}{Zhou Zhao}.} \bibinfo{year}{2024}\natexlab{}.
\newblock \showarticletitle{WavTokenizer: an Efficient Acoustic Discrete Codec Tokenizer for Audio Language Modeling}.
\newblock \bibinfo{journal}{\emph{CoRR}}  \bibinfo{volume}{abs/2408.16532} (\bibinfo{year}{2024}).
\newblock


\bibitem[Jiang et~al\mbox{.}(2023)]%
        {megatts2}
\bibfield{author}{\bibinfo{person}{Ziyue Jiang}, \bibinfo{person}{Jinglin Liu}, \bibinfo{person}{Yi Ren}, \bibinfo{person}{Jinzheng He}, \bibinfo{person}{Chen Zhang}, \bibinfo{person}{Zhenhui Ye}, \bibinfo{person}{Pengfei Wei}, \bibinfo{person}{Chunfeng Wang}, \bibinfo{person}{Xiang Yin}, \bibinfo{person}{Zejun Ma}, {and} \bibinfo{person}{Zhou Zhao}.} \bibinfo{year}{2023}\natexlab{}.
\newblock \showarticletitle{Mega-TTS 2: Zero-Shot Text-to-Speech with Arbitrary Length Speech Prompts}.
\newblock \bibinfo{journal}{\emph{CoRR}}  \bibinfo{volume}{abs/2307.07218} (\bibinfo{year}{2023}).
\newblock


\bibitem[Ju et~al\mbox{.}(2024)]%
        {ns3}
\bibfield{author}{\bibinfo{person}{Zeqian Ju}, \bibinfo{person}{Yuancheng Wang}, \bibinfo{person}{Kai Shen}, \bibinfo{person}{Xu Tan}, \bibinfo{person}{Detai Xin}, \bibinfo{person}{Dongchao Yang}, \bibinfo{person}{Eric Liu}, \bibinfo{person}{Yichong Leng}, \bibinfo{person}{Kaitao Song}, \bibinfo{person}{Siliang Tang}, \bibinfo{person}{Zhizheng Wu}, \bibinfo{person}{Tao Qin}, \bibinfo{person}{Xiangyang Li}, \bibinfo{person}{Wei Ye}, \bibinfo{person}{Shikun Zhang}, \bibinfo{person}{Jiang Bian}, \bibinfo{person}{Lei He}, \bibinfo{person}{Jinyu Li}, {and} \bibinfo{person}{Sheng Zhao}.} \bibinfo{year}{2024}\natexlab{}.
\newblock \showarticletitle{NaturalSpeech 3: Zero-Shot Speech Synthesis with Factorized Codec and Diffusion Models}. In \bibinfo{booktitle}{\emph{Forty-first International Conference on Machine Learning, {ICML} 2024, Vienna, Austria, July 21-27, 2024}}. \bibinfo{publisher}{OpenReview.net}.
\newblock


\bibitem[Kilgour et~al\mbox{.}(2018)]%
        {fad}
\bibfield{author}{\bibinfo{person}{Kevin Kilgour}, \bibinfo{person}{Mauricio Zuluaga}, \bibinfo{person}{Dominik Roblek}, {and} \bibinfo{person}{Matthew Sharifi}.} \bibinfo{year}{2018}\natexlab{}.
\newblock \showarticletitle{Fr$\backslash$'echet audio distance: A metric for evaluating music enhancement algorithms}.
\newblock \bibinfo{journal}{\emph{arXiv preprint arXiv:1812.08466}} (\bibinfo{year}{2018}).
\newblock


\bibitem[Kong et~al\mbox{.}(2020)]%
        {panns}
\bibfield{author}{\bibinfo{person}{Qiuqiang Kong}, \bibinfo{person}{Yin Cao}, \bibinfo{person}{Turab Iqbal}, \bibinfo{person}{Yuxuan Wang}, \bibinfo{person}{Wenwu Wang}, {and} \bibinfo{person}{Mark~D Plumbley}.} \bibinfo{year}{2020}\natexlab{}.
\newblock \showarticletitle{Panns: Large-scale pretrained audio neural networks for audio pattern recognition}.
\newblock \bibinfo{journal}{\emph{IEEE/ACM Transactions on Audio, Speech, and Language Processing}}  \bibinfo{volume}{28} (\bibinfo{year}{2020}), \bibinfo{pages}{2880--2894}.
\newblock


\bibitem[Kumar et~al\mbox{.}(2019)]%
        {Lipper}
\bibfield{author}{\bibinfo{person}{Yaman Kumar}, \bibinfo{person}{Rohit Jain}, \bibinfo{person}{Khwaja~Mohd. Salik}, \bibinfo{person}{Rajiv~Ratn Shah}, \bibinfo{person}{Yifang Yin}, {and} \bibinfo{person}{Roger Zimmermann}.} \bibinfo{year}{2019}\natexlab{}.
\newblock \showarticletitle{Lipper: Synthesizing Thy Speech Using Multi-View Lipreading}. In \bibinfo{booktitle}{\emph{The Thirty-Third {AAAI} Conference on Artificial Intelligence, {AAAI} 2019, The Thirty-First Innovative Applications of Artificial Intelligence Conference, {IAAI} 2019, The Ninth {AAAI} Symposium on Educational Advances in Artificial Intelligence, {EAAI} 2019, Honolulu, Hawaii, USA, January 27 - February 1, 2019}}. \bibinfo{publisher}{{AAAI} Press}, \bibinfo{pages}{2588--2595}.
\newblock


\bibitem[Lee et~al\mbox{.}(2023)]%
        {facetts}
\bibfield{author}{\bibinfo{person}{Jiyoung Lee}, \bibinfo{person}{Joon~Son Chung}, {and} \bibinfo{person}{Soo{-}Whan Chung}.} \bibinfo{year}{2023}\natexlab{}.
\newblock \showarticletitle{Imaginary Voice: Face-Styled Diffusion Model for Text-to-Speech}. In \bibinfo{booktitle}{\emph{{IEEE} International Conference on Acoustics, Speech and Signal Processing {ICASSP} 2023, Rhodes Island, Greece, June 4-10, 2023}}. \bibinfo{publisher}{{IEEE}}, \bibinfo{pages}{1--5}.
\newblock


\bibitem[Lei et~al\mbox{.}(2024)]%
        {aligner}
\bibfield{author}{\bibinfo{person}{Shun Lei}, \bibinfo{person}{Yixuan Zhou}, \bibinfo{person}{Boshi Tang}, \bibinfo{person}{Max~WY Lam}, \bibinfo{person}{Hangyu Liu}, \bibinfo{person}{Jingcheng Wu}, \bibinfo{person}{Shiyin Kang}, \bibinfo{person}{Zhiyong Wu}, \bibinfo{person}{Helen Meng}, {et~al\mbox{.}}} \bibinfo{year}{2024}\natexlab{}.
\newblock \showarticletitle{Songcreator: Lyrics-based universal song generation}.
\newblock \bibinfo{journal}{\emph{Advances in Neural Information Processing Systems}}  \bibinfo{volume}{37} (\bibinfo{year}{2024}), \bibinfo{pages}{80107--80140}.
\newblock


\bibitem[Li et~al\mbox{.}(2023)]%
        {styletts2}
\bibfield{author}{\bibinfo{person}{Yinghao~Aaron Li}, \bibinfo{person}{Cong Han}, \bibinfo{person}{Vinay~S. Raghavan}, \bibinfo{person}{Gavin Mischler}, {and} \bibinfo{person}{Nima Mesgarani}.} \bibinfo{year}{2023}\natexlab{}.
\newblock \showarticletitle{StyleTTS 2: Towards Human-Level Text-to-Speech through Style Diffusion and Adversarial Training with Large Speech Language Models}. In \bibinfo{booktitle}{\emph{Advances in Neural Information Processing Systems 36: Annual Conference on Neural Information Processing Systems 2023, NeurIPS 2023, New Orleans, LA, USA, December 10 - 16, 2023}}, \bibfield{editor}{\bibinfo{person}{Alice Oh}, \bibinfo{person}{Tristan Naumann}, \bibinfo{person}{Amir Globerson}, \bibinfo{person}{Kate Saenko}, \bibinfo{person}{Moritz Hardt}, {and} \bibinfo{person}{Sergey Levine}} (Eds.).
\newblock


\bibitem[Liu et~al\mbox{.}(2023)]%
        {audioldm}
\bibfield{author}{\bibinfo{person}{Haohe Liu}, \bibinfo{person}{Zehua Chen}, \bibinfo{person}{Yi Yuan}, \bibinfo{person}{Xinhao Mei}, \bibinfo{person}{Xubo Liu}, \bibinfo{person}{Danilo~P. Mandic}, \bibinfo{person}{Wenwu Wang}, {and} \bibinfo{person}{Mark~D. Plumbley}.} \bibinfo{year}{2023}\natexlab{}.
\newblock \showarticletitle{AudioLDM: Text-to-Audio Generation with Latent Diffusion Models}. In \bibinfo{booktitle}{\emph{International Conference on Machine Learning, {ICML} 2023, 23-29 July 2023, Honolulu, Hawaii, {USA}}} \emph{(\bibinfo{series}{Proceedings of Machine Learning Research}, Vol.~\bibinfo{volume}{202})}, \bibfield{editor}{\bibinfo{person}{Andreas Krause}, \bibinfo{person}{Emma Brunskill}, \bibinfo{person}{Kyunghyun Cho}, \bibinfo{person}{Barbara Engelhardt}, \bibinfo{person}{Sivan Sabato}, {and} \bibinfo{person}{Jonathan Scarlett}} (Eds.). \bibinfo{publisher}{{PMLR}}, \bibinfo{pages}{21450--21474}.
\newblock


\bibitem[Luo et~al\mbox{.}(2023)]%
        {Diff-foley}
\bibfield{author}{\bibinfo{person}{Simian Luo}, \bibinfo{person}{Chuanhao Yan}, \bibinfo{person}{Chenxu Hu}, {and} \bibinfo{person}{Hang Zhao}.} \bibinfo{year}{2023}\natexlab{}.
\newblock \showarticletitle{Diff-foley: Synchronized video-to-audio synthesis with latent diffusion models}.
\newblock \bibinfo{journal}{\emph{Advances in Neural Information Processing Systems}}  \bibinfo{volume}{36} (\bibinfo{year}{2023}), \bibinfo{pages}{48855--48876}.
\newblock


\bibitem[Meng et~al\mbox{.}(2024)]%
        {melle}
\bibfield{author}{\bibinfo{person}{Lingwei Meng}, \bibinfo{person}{Long Zhou}, \bibinfo{person}{Shujie Liu}, \bibinfo{person}{Sanyuan Chen}, \bibinfo{person}{Bing Han}, \bibinfo{person}{Shujie Hu}, \bibinfo{person}{Yanqing Liu}, \bibinfo{person}{Jinyu Li}, \bibinfo{person}{Sheng Zhao}, \bibinfo{person}{Xixin Wu}, {et~al\mbox{.}}} \bibinfo{year}{2024}\natexlab{}.
\newblock \showarticletitle{Autoregressive speech synthesis without vector quantization}.
\newblock \bibinfo{journal}{\emph{arXiv preprint arXiv:2407.08551}} (\bibinfo{year}{2024}).
\newblock


\bibitem[Radford et~al\mbox{.}(2021)]%
        {clip}
\bibfield{author}{\bibinfo{person}{Alec Radford}, \bibinfo{person}{Jong~Wook Kim}, \bibinfo{person}{Chris Hallacy}, \bibinfo{person}{Aditya Ramesh}, \bibinfo{person}{Gabriel Goh}, \bibinfo{person}{Sandhini Agarwal}, \bibinfo{person}{Girish Sastry}, \bibinfo{person}{Amanda Askell}, \bibinfo{person}{Pamela Mishkin}, \bibinfo{person}{Jack Clark}, \bibinfo{person}{Gretchen Krueger}, {and} \bibinfo{person}{Ilya Sutskever}.} \bibinfo{year}{2021}\natexlab{}.
\newblock \showarticletitle{Learning Transferable Visual Models From Natural Language Supervision}. In \bibinfo{booktitle}{\emph{Proceedings of the 38th International Conference on Machine Learning, {ICML} 2021, 18-24 July 2021, Virtual Event}} \emph{(\bibinfo{series}{Proceedings of Machine Learning Research}, Vol.~\bibinfo{volume}{139})}, \bibfield{editor}{\bibinfo{person}{Marina Meila} {and} \bibinfo{person}{Tong Zhang}} (Eds.). \bibinfo{publisher}{{PMLR}}, \bibinfo{pages}{8748--8763}.
\newblock


\bibitem[Radford et~al\mbox{.}(2023)]%
        {whisper}
\bibfield{author}{\bibinfo{person}{Alec Radford}, \bibinfo{person}{Jong~Wook Kim}, \bibinfo{person}{Tao Xu}, \bibinfo{person}{Greg Brockman}, \bibinfo{person}{Christine McLeavey}, {and} \bibinfo{person}{Ilya Sutskever}.} \bibinfo{year}{2023}\natexlab{}.
\newblock \showarticletitle{Robust speech recognition via large-scale weak supervision}. In \bibinfo{booktitle}{\emph{International conference on machine learning}}. PMLR, \bibinfo{pages}{28492--28518}.
\newblock


\bibitem[Saeki et~al\mbox{.}(2022)]%
        {utmos}
\bibfield{author}{\bibinfo{person}{Takaaki Saeki}, \bibinfo{person}{Detai Xin}, \bibinfo{person}{Wataru Nakata}, \bibinfo{person}{Tomoki Koriyama}, \bibinfo{person}{Shinnosuke Takamichi}, {and} \bibinfo{person}{Hiroshi Saruwatari}.} \bibinfo{year}{2022}\natexlab{}.
\newblock \showarticletitle{Utmos: Utokyo-sarulab system for voicemos challenge 2022}.
\newblock \bibinfo{journal}{\emph{arXiv preprint arXiv:2204.02152}} (\bibinfo{year}{2022}).
\newblock


\bibitem[Sheffer and Adi(2023)]%
        {im2wav}
\bibfield{author}{\bibinfo{person}{Roy Sheffer} {and} \bibinfo{person}{Yossi Adi}.} \bibinfo{year}{2023}\natexlab{}.
\newblock \showarticletitle{I hear your true colors: Image guided audio generation}. In \bibinfo{booktitle}{\emph{ICASSP 2023-2023 IEEE International Conference on Acoustics, Speech and Signal Processing (ICASSP)}}. IEEE, \bibinfo{pages}{1--5}.
\newblock


\bibitem[Shen et~al\mbox{.}(2024)]%
        {ns2}
\bibfield{author}{\bibinfo{person}{Kai Shen}, \bibinfo{person}{Zeqian Ju}, \bibinfo{person}{Xu Tan}, \bibinfo{person}{Eric Liu}, \bibinfo{person}{Yichong Leng}, \bibinfo{person}{Lei He}, \bibinfo{person}{Tao Qin}, \bibinfo{person}{Sheng Zhao}, {and} \bibinfo{person}{Jiang Bian}.} \bibinfo{year}{2024}\natexlab{}.
\newblock \showarticletitle{NaturalSpeech 2: Latent Diffusion Models are Natural and Zero-Shot Speech and Singing Synthesizers}. In \bibinfo{booktitle}{\emph{The Twelfth International Conference on Learning Representations, {ICLR} 2024, Vienna, Austria, May 7-11, 2024}}. \bibinfo{publisher}{OpenReview.net}.
\newblock


\bibitem[Tang et~al\mbox{.}(2024)]%
        {lm3}
\bibfield{author}{\bibinfo{person}{Zineng Tang}, \bibinfo{person}{Ziyi Yang}, \bibinfo{person}{Mahmoud Khademi}, \bibinfo{person}{Yang Liu}, \bibinfo{person}{Chenguang Zhu}, {and} \bibinfo{person}{Mohit Bansal}.} \bibinfo{year}{2024}\natexlab{}.
\newblock \showarticletitle{CoDi-2: In-Context, Interleaved, and Interactive Any-to-Any Generation}. In \bibinfo{booktitle}{\emph{{IEEE/CVF} Conference on Computer Vision and Pattern Recognition, {CVPR} 2024, Seattle, WA, USA, June 16-22, 2024}}. \bibinfo{publisher}{{IEEE}}, \bibinfo{pages}{27415--27424}.
\newblock


\bibitem[Touvron et~al\mbox{.}(2023)]%
        {llama}
\bibfield{author}{\bibinfo{person}{Hugo Touvron}, \bibinfo{person}{Thibaut Lavril}, \bibinfo{person}{Gautier Izacard}, \bibinfo{person}{Xavier Martinet}, \bibinfo{person}{Marie{-}Anne Lachaux}, \bibinfo{person}{Timoth{\'{e}}e Lacroix}, \bibinfo{person}{Baptiste Rozi{\`{e}}re}, \bibinfo{person}{Naman Goyal}, \bibinfo{person}{Eric Hambro}, \bibinfo{person}{Faisal Azhar}, \bibinfo{person}{Aur{\'{e}}lien Rodriguez}, \bibinfo{person}{Armand Joulin}, \bibinfo{person}{Edouard Grave}, {and} \bibinfo{person}{Guillaume Lample}.} \bibinfo{year}{2023}\natexlab{}.
\newblock \showarticletitle{LLaMA: Open and Efficient Foundation Language Models}.
\newblock \bibinfo{journal}{\emph{CoRR}}  \bibinfo{volume}{abs/2302.13971} (\bibinfo{year}{2023}).
\newblock


\bibitem[Wang et~al\mbox{.}(2025)]%
        {FELLE}
\bibfield{author}{\bibinfo{person}{Hui Wang}, \bibinfo{person}{Shujie Liu}, \bibinfo{person}{Lingwei Meng}, \bibinfo{person}{Jinyu Li}, \bibinfo{person}{Yifan Yang}, \bibinfo{person}{Shiwan Zhao}, \bibinfo{person}{Haiyang Sun}, \bibinfo{person}{Yanqing Liu}, \bibinfo{person}{Haoqin Sun}, \bibinfo{person}{Jiaming Zhou}, \bibinfo{person}{Yan Lu}, {and} \bibinfo{person}{Yong Qin}.} \bibinfo{year}{2025}\natexlab{}.
\newblock \showarticletitle{{FELLE:} Autoregressive Speech Synthesis with Token-Wise Coarse-to-Fine Flow Matching}.
\newblock \bibinfo{journal}{\emph{CoRR}}  \bibinfo{volume}{abs/2502.11128} (\bibinfo{year}{2025}).
\newblock


\bibitem[Wang et~al\mbox{.}(2024b)]%
        {V2a-mapper}
\bibfield{author}{\bibinfo{person}{Heng Wang}, \bibinfo{person}{Jianbo Ma}, \bibinfo{person}{Santiago Pascual}, \bibinfo{person}{Richard Cartwright}, {and} \bibinfo{person}{Weidong Cai}.} \bibinfo{year}{2024}\natexlab{b}.
\newblock \showarticletitle{V2A-Mapper: {A} Lightweight Solution for Vision-to-Audio Generation by Connecting Foundation Models}. In \bibinfo{booktitle}{\emph{Thirty-Eighth {AAAI} Conference on Artificial Intelligence, {AAAI} 2024, Thirty-Sixth Conference on Innovative Applications of Artificial Intelligence, {IAAI} 2024, Fourteenth Symposium on Educational Advances in Artificial Intelligence, {EAAI} 2014, February 20-27, 2024, Vancouver, Canada}}, \bibfield{editor}{\bibinfo{person}{Michael~J. Wooldridge}, \bibinfo{person}{Jennifer~G. Dy}, {and} \bibinfo{person}{Sriraam Natarajan}} (Eds.). \bibinfo{publisher}{{AAAI} Press}, \bibinfo{pages}{15492--15501}.
\newblock


\bibitem[Wang et~al\mbox{.}(2023)]%
        {melroformer}
\bibfield{author}{\bibinfo{person}{Ju-Chiang Wang}, \bibinfo{person}{Wei-Tsung Lu}, {and} \bibinfo{person}{Minz Won}.} \bibinfo{year}{2023}\natexlab{}.
\newblock \showarticletitle{Mel-Band RoFormer for Music Source Separation}.
\newblock \bibinfo{journal}{\emph{arXiv preprint arXiv:2310.01809}} (\bibinfo{year}{2023}).
\newblock


\bibitem[Wang et~al\mbox{.}(2024a)]%
        {Frieren}
\bibfield{author}{\bibinfo{person}{Yongqi Wang}, \bibinfo{person}{Wenxiang Guo}, \bibinfo{person}{Rongjie Huang}, \bibinfo{person}{Jiawei Huang}, \bibinfo{person}{Zehan Wang}, \bibinfo{person}{Fuming You}, \bibinfo{person}{Ruiqi Li}, {and} \bibinfo{person}{Zhou Zhao}.} \bibinfo{year}{2024}\natexlab{a}.
\newblock \showarticletitle{Frieren: Efficient Video-to-Audio Generation Network with Rectified Flow Matching}. In \bibinfo{booktitle}{\emph{Advances in Neural Information Processing Systems 38: Annual Conference on Neural Information Processing Systems 2024, NeurIPS 2024, Vancouver, BC, Canada, December 10 - 15, 2024}}, \bibfield{editor}{\bibinfo{person}{Amir Globersons}, \bibinfo{person}{Lester Mackey}, \bibinfo{person}{Danielle Belgrave}, \bibinfo{person}{Angela Fan}, \bibinfo{person}{Ulrich Paquet}, \bibinfo{person}{Jakub~M. Tomczak}, {and} \bibinfo{person}{Cheng Zhang}} (Eds.).
\newblock


\bibitem[Xie et~al\mbox{.}(2024)]%
        {Sonicvisionlm}
\bibfield{author}{\bibinfo{person}{Zhifeng Xie}, \bibinfo{person}{Shengye Yu}, \bibinfo{person}{Qile He}, {and} \bibinfo{person}{Mengtian Li}.} \bibinfo{year}{2024}\natexlab{}.
\newblock \showarticletitle{Sonicvisionlm: Playing sound with vision language models}. In \bibinfo{booktitle}{\emph{Proceedings of the IEEE/CVF Conference on Computer Vision and Pattern Recognition}}. \bibinfo{pages}{26866--26875}.
\newblock


\bibitem[Xing et~al\mbox{.}(2024)]%
        {seeandhear}
\bibfield{author}{\bibinfo{person}{Yazhou Xing}, \bibinfo{person}{Yingqing He}, \bibinfo{person}{Zeyue Tian}, \bibinfo{person}{Xintao Wang}, {and} \bibinfo{person}{Qifeng Chen}.} \bibinfo{year}{2024}\natexlab{}.
\newblock \showarticletitle{Seeing and Hearing: Open-domain Visual-Audio Generation with Diffusion Latent Aligners}. In \bibinfo{booktitle}{\emph{{IEEE/CVF} Conference on Computer Vision and Pattern Recognition, {CVPR} 2024, Seattle, WA, USA, June 16-22, 2024}}. \bibinfo{publisher}{{IEEE}}, \bibinfo{pages}{7151--7161}.
\newblock


\bibitem[Yang et~al\mbox{.}(2024)]%
        {qwen2}
\bibfield{author}{\bibinfo{person}{An Yang}, \bibinfo{person}{Baosong Yang}, \bibinfo{person}{Binyuan Hui}, \bibinfo{person}{Bo Zheng}, \bibinfo{person}{Bowen Yu}, \bibinfo{person}{Chang Zhou}, \bibinfo{person}{Chengpeng Li}, \bibinfo{person}{Chengyuan Li}, \bibinfo{person}{Dayiheng Liu}, \bibinfo{person}{Fei Huang}, \bibinfo{person}{Guanting Dong}, \bibinfo{person}{Haoran Wei}, \bibinfo{person}{Huan Lin}, \bibinfo{person}{Jialong Tang}, \bibinfo{person}{Jialin Wang}, \bibinfo{person}{Jian Yang}, \bibinfo{person}{Jianhong Tu}, \bibinfo{person}{Jianwei Zhang}, \bibinfo{person}{Jianxin Ma}, \bibinfo{person}{Jianxin Yang}, \bibinfo{person}{Jin Xu}, \bibinfo{person}{Jingren Zhou}, \bibinfo{person}{Jinze Bai}, \bibinfo{person}{Jinzheng He}, \bibinfo{person}{Junyang Lin}, \bibinfo{person}{Kai Dang}, \bibinfo{person}{Keming Lu}, \bibinfo{person}{Keqin Chen}, \bibinfo{person}{Kexin Yang}, \bibinfo{person}{Mei Li}, \bibinfo{person}{Mingfeng Xue}, \bibinfo{person}{Na Ni}, \bibinfo{person}{Pei Zhang},
  \bibinfo{person}{Peng Wang}, \bibinfo{person}{Ru Peng}, \bibinfo{person}{Rui Men}, \bibinfo{person}{Ruize Gao}, \bibinfo{person}{Runji Lin}, \bibinfo{person}{Shijie Wang}, \bibinfo{person}{Shuai Bai}, \bibinfo{person}{Sinan Tan}, \bibinfo{person}{Tianhang Zhu}, \bibinfo{person}{Tianhao Li}, \bibinfo{person}{Tianyu Liu}, \bibinfo{person}{Wenbin Ge}, \bibinfo{person}{Xiaodong Deng}, \bibinfo{person}{Xiaohuan Zhou}, \bibinfo{person}{Xingzhang Ren}, \bibinfo{person}{Xinyu Zhang}, \bibinfo{person}{Xipin Wei}, \bibinfo{person}{Xuancheng Ren}, \bibinfo{person}{Xuejing Liu}, \bibinfo{person}{Yang Fan}, \bibinfo{person}{Yang Yao}, \bibinfo{person}{Yichang Zhang}, \bibinfo{person}{Yu Wan}, \bibinfo{person}{Yunfei Chu}, \bibinfo{person}{Yuqiong Liu}, \bibinfo{person}{Zeyu Cui}, \bibinfo{person}{Zhenru Zhang}, \bibinfo{person}{Zhifang Guo}, {and} \bibinfo{person}{Zhihao Fan}.} \bibinfo{year}{2024}\natexlab{}.
\newblock \showarticletitle{Qwen2 Technical Report}.
\newblock \bibinfo{journal}{\emph{CoRR}}  \bibinfo{volume}{abs/2407.10671} (\bibinfo{year}{2024}).
\newblock


\bibitem[Yang et~al\mbox{.}(2023)]%
        {uniaudiolm}
\bibfield{author}{\bibinfo{person}{Dongchao Yang}, \bibinfo{person}{Jinchuan Tian}, \bibinfo{person}{Xu Tan}, \bibinfo{person}{Rongjie Huang}, \bibinfo{person}{Songxiang Liu}, \bibinfo{person}{Xuankai Chang}, \bibinfo{person}{Jiatong Shi}, \bibinfo{person}{Sheng Zhao}, \bibinfo{person}{Jiang Bian}, \bibinfo{person}{Xixin Wu}, \bibinfo{person}{Zhou Zhao}, \bibinfo{person}{Shinji Watanabe}, {and} \bibinfo{person}{Helen Meng}.} \bibinfo{year}{2023}\natexlab{}.
\newblock \showarticletitle{UniAudio: An Audio Foundation Model Toward Universal Audio Generation}.
\newblock \bibinfo{journal}{\emph{CoRR}}  \bibinfo{volume}{abs/2310.00704} (\bibinfo{year}{2023}).
\newblock


\bibitem[Yariv et~al\mbox{.}(2023)]%
        {AValigntool}
\bibfield{author}{\bibinfo{person}{Guy Yariv}, \bibinfo{person}{Itai Gat}, \bibinfo{person}{Sagie Benaim}, \bibinfo{person}{Lior Wolf}, \bibinfo{person}{Idan Schwartz}, {and} \bibinfo{person}{Yossi Adi}.} \bibinfo{year}{2023}\natexlab{}.
\newblock \bibinfo{title}{Diverse and Aligned Audio-to-Video Generation via Text-to-Video Model Adaptation}.
\newblock
\newblock
\showeprint[arxiv]{2309.16429}~[cs.LG]


\bibitem[Ye et~al\mbox{.}(2025)]%
        {llasa}
\bibfield{author}{\bibinfo{person}{Zhen Ye}, \bibinfo{person}{Xinfa Zhu}, \bibinfo{person}{Chi-Min Chan}, \bibinfo{person}{Xinsheng Wang}, \bibinfo{person}{Xu Tan}, \bibinfo{person}{Jiahe Lei}, \bibinfo{person}{Yi Peng}, \bibinfo{person}{Haohe Liu}, \bibinfo{person}{Yizhu Jin}, \bibinfo{person}{Zheqi DAI}, {et~al\mbox{.}}} \bibinfo{year}{2025}\natexlab{}.
\newblock \showarticletitle{Llasa: Scaling Train-Time and Inference-Time Compute for Llama-based Speech Synthesis}.
\newblock \bibinfo{journal}{\emph{arXiv preprint arXiv:2502.04128}} (\bibinfo{year}{2025}).
\newblock


\bibitem[Zen et~al\mbox{.}(2019)]%
        {librittsdata}
\bibfield{author}{\bibinfo{person}{Heiga Zen}, \bibinfo{person}{Viet Dang}, \bibinfo{person}{Rob Clark}, \bibinfo{person}{Yu Zhang}, \bibinfo{person}{Ron~J Weiss}, \bibinfo{person}{Ye Jia}, \bibinfo{person}{Zhifeng Chen}, {and} \bibinfo{person}{Yonghui Wu}.} \bibinfo{year}{2019}\natexlab{}.
\newblock \showarticletitle{Libritts: A corpus derived from librispeech for text-to-speech}.
\newblock \bibinfo{journal}{\emph{arXiv preprint arXiv:1904.02882}} (\bibinfo{year}{2019}).
\newblock


\bibitem[Zhang et~al\mbox{.}(2024a)]%
        {FoleyCrafter}
\bibfield{author}{\bibinfo{person}{Yiming Zhang}, \bibinfo{person}{Yicheng Gu}, \bibinfo{person}{Yanhong Zeng}, \bibinfo{person}{Zhening Xing}, \bibinfo{person}{Yuancheng Wang}, \bibinfo{person}{Zhizheng Wu}, {and} \bibinfo{person}{Kai Chen}.} \bibinfo{year}{2024}\natexlab{a}.
\newblock \showarticletitle{FoleyCrafter: Bring Silent Videos to Life with Lifelike and Synchronized Sounds}.
\newblock \bibinfo{journal}{\emph{CoRR}}  \bibinfo{volume}{abs/2407.01494} (\bibinfo{year}{2024}).
\newblock


\bibitem[Zhang et~al\mbox{.}(2024b)]%
        {speechdubber}
\bibfield{author}{\bibinfo{person}{Zhedong Zhang}, \bibinfo{person}{Liang Li}, \bibinfo{person}{Gaoxiang Cong}, \bibinfo{person}{Haibing Yin}, \bibinfo{person}{Yuhan Gao}, \bibinfo{person}{Chenggang Yan}, \bibinfo{person}{Anton van~den Hengel}, {and} \bibinfo{person}{Yuankai Qi}.} \bibinfo{year}{2024}\natexlab{b}.
\newblock \showarticletitle{From Speaker to Dubber: Movie Dubbing with Prosody and Duration Consistency Learning}. In \bibinfo{booktitle}{\emph{Proceedings of the 32nd {ACM} International Conference on Multimedia, {MM} 2024, Melbourne, VIC, Australia, 28 October 2024 - 1 November 2024}}, \bibfield{editor}{\bibinfo{person}{Jianfei Cai}, \bibinfo{person}{Mohan~S. Kankanhalli}, \bibinfo{person}{Balakrishnan Prabhakaran}, \bibinfo{person}{Susanne Boll}, \bibinfo{person}{Ramanathan Subramanian}, \bibinfo{person}{Liang Zheng}, \bibinfo{person}{Vivek~K. Singh}, \bibinfo{person}{Pablo C{\'{e}}sar}, \bibinfo{person}{Lexing Xie}, {and} \bibinfo{person}{Dong Xu}} (Eds.). \bibinfo{publisher}{{ACM}}, \bibinfo{pages}{7523--7532}.
\newblock


\bibitem[Zhao et~al\mbox{.}(2024)]%
        {mcdubber}
\bibfield{author}{\bibinfo{person}{Yuan Zhao}, \bibinfo{person}{Zhenqi Jia}, \bibinfo{person}{Rui Liu}, \bibinfo{person}{De Hu}, \bibinfo{person}{Feilong Bao}, {and} \bibinfo{person}{Guanglai Gao}.} \bibinfo{year}{2024}\natexlab{}.
\newblock \showarticletitle{Mcdubber: Multimodal context-aware expressive video dubbing}. In \bibinfo{booktitle}{\emph{National Conference on Man-Machine Speech Communication}}. Springer, \bibinfo{pages}{168--182}.
\newblock


\bibitem[Zhu et~al\mbox{.}(2025)]%
        {cosyaudio}
\bibfield{author}{\bibinfo{person}{Xinfa Zhu}, \bibinfo{person}{Wenjie Tian}, \bibinfo{person}{Xinsheng Wang}, \bibinfo{person}{Lei He}, \bibinfo{person}{Xi Wang}, \bibinfo{person}{Sheng Zhao}, {and} \bibinfo{person}{Lei Xie}.} \bibinfo{year}{2025}\natexlab{}.
\newblock \showarticletitle{CosyAudio: Improving Audio Generation with Confidence Scores and Synthetic Captions}.
\newblock \bibinfo{journal}{\emph{CoRR}}  \bibinfo{volume}{abs/2501.16761} (\bibinfo{year}{2025}).
\newblock


\bibitem[Zhu et~al\mbox{.}(2024)]%
        {kalle}
\bibfield{author}{\bibinfo{person}{Xinfa Zhu}, \bibinfo{person}{Wenjie Tian}, {and} \bibinfo{person}{Lei Xie}.} \bibinfo{year}{2024}\natexlab{}.
\newblock \showarticletitle{Autoregressive Speech Synthesis with Next-Distribution Prediction}.
\newblock \bibinfo{journal}{\emph{arXiv preprint arXiv:2412.16846}} (\bibinfo{year}{2024}).
\newblock


\end{thebibliography}

\end{document}